\documentclass[twocolumn]{autart}    

\usepackage{color}
\usepackage{graphics} 
\usepackage{epsfig} 
\usepackage{mathptmx} 
\usepackage{times} 
\usepackage{amsmath} 
\usepackage{amssymb}  
\usepackage{upgreek}
\usepackage{latexsym,,graphicx,epsf,cite,url,bbm,float,subfig,amsbsy}
\usepackage{ifpdf}
\usepackage{epstopdf}
\usepackage{color}
\usepackage{tabularx}
\usepackage{stfloats}
\usepackage{tikz}
 
\usepackage[percent]{overpic}
\usepackage[utf8]{inputenc}

\makeatletter
\DeclareFontFamily{U}  {MnSymbolF}{}
\DeclareSymbolFont{symbolsMN}{U}{MnSymbolF}{m}{n}
\SetSymbolFont{symbolsMN}{bold}{U}{MnSymbolF}{b}{n}
\DeclareFontShape{U}{MnSymbolF}{m}{n}{
    <-6>  MnSymbolF5
   <6-7>  MnSymbolF6
   <7-8>  MnSymbolF7
   <8-9>  MnSymbolF8
   <9-10> MnSymbolF9
  <10-12> MnSymbolF10
  <12->   MnSymbolF12}{}
\DeclareFontShape{U}{MnSymbolF}{b}{n}{
    <-6>  MnSymbolF-Bold5
   <6-7>  MnSymbolF-Bold6
   <7-8>  MnSymbolF-Bold7
   <8-9>  MnSymbolF-Bold8
   <9-10> MnSymbolF-Bold9
  <10-12> MnSymbolF-Bold10
  <12->   MnSymbolF-Bold12}{}
\DeclareMathSymbol{\tbigtimes}{\mathop}{symbolsMN}{2}
\newcommand*{\bigtimes}{%
  \DOTSB
  \tbigtimes
  \slimits@
}
\makeatother

\newcommand{\boxalign}[2][0.97\textwidth]{
  \par\noindent\tikzstyle{mybox} = [draw=black,inner sep=6pt]
  \begin{center}\begin{tikzpicture}
   \node [mybox] (box){%
    \begin{minipage}{#1}{\vspace{-5mm}#2}\end{minipage}
   };
  \end{tikzpicture}\end{center}
}

\newcommand{\cb}{}

\newcommand{\nn}{\nonumber}
\newcommand{\E}{\mathbb{E}}
\newcommand{\rx}{\pmb{x}}
\newcommand{\rtheta}{\pmb{\theta}}
\newcommand{\romega}{\pmb{\omega}}
\newcommand{\rnu}{\pmb{\nu}}
\newcommand{\rw}{\pmb{w}}
\newcommand{\rv}{\pmb{v}}
\newcommand{\rs}{\pmb{z}}

\newcommand{\ry}{\pmb{y}}
\newcommand{\ru}{\pmb{u}}

\newcommand{\rz}{\pmb{z}}
\newcommand{\hx}{\pmb{\hat{x}}}

\newcommand{\React}{\Pi}
\newcommand{\Q}{Q}
\newcommand{\R}{R}
\renewcommand{\H}{H}
\newcommand{\B}{L}
\newcommand{\BB}{K}
\newcommand{\C}{\mathcal{C}}
\newcommand{\K}{K}
\newcommand{\tQ}{\tilde{\Q}}

\newcommand{\signal}{\eta^{\ell}}
\newcommand{\ryl}{\ry^{\ell}}

\newenvironment{psmallmatrix}
  {\left[\begin{smallmatrix}}
  {\end{smallmatrix}\right]}

\begin{document}

\begin{frontmatter}

\title{Hierarchical Multistage Gaussian Signaling Games {\cb in Noncooperative Communication and Control Systems}\thanksref{footnoteinfo}} 

\thanks[footnoteinfo]{This paper was not presented at any IFAC meeting. Corresponding author M.~O.~Sayin. Tel. +1-217-898-5476.\\
\textit{Email addresses:} sayin2@illinois.edu (Muhammed O. Sayin), eakyol@binghamton.edu (Emrah Akyol),  basar1@illinois.edu (Tamer Ba\c{s}ar).}



\author[UIUC]{Muhammed O. Sayin},~ 
\author[SUNY]{Emrah Akyol},~and~
\author[UIUC]{Tamer Ba\c{s}ar}

\address[UIUC]{Coordinated Science Laboratory, University of Illinois at Urbana-Champaign, Urbana, IL 61801}  
\address[SUNY]{Binghamton University, State University of New York, Binghamton, NY 13902}             


\begin{abstract}                          
We analyze in this paper finite horizon hierarchical signaling games between (information provider) senders and (decision maker) receivers in a dynamic environment. The underlying information evolves in time while sender and receiver interact repeatedly. Different from the classical communication (control) models, however, the sender (sensor) and the receiver (controller) have different objectives and there is a hierarchy between the players such that the sender leads the game by announcing his policies beforehand. He needs to anticipate the reaction of the receiver and the impact of the actions on the horizon while controlling the transparency of the disclosed information at each interaction. With quadratic cost functions and multivariate Gaussian processes, evolving according to first order auto-regressive models, we show that memoryless ``linear" sender signaling rules are optimal (in the sense of game-theoretic hierarchical equilibrium) within the general class of measurable policies in the noncooperative communication context. {\cb In the noncooperative control context, we also analyze the hierarchical equilibrium for linear signaling rules and provide an algorithm to compute the optimal linear signaling rules numerically with global optimality guarantees.}
\end{abstract}

\end{frontmatter}

\section{Introduction}
In the era of smart devices, we have various systems having enhanced processing and efficient communication capabilities. Even though information exchange is generally useful in cooperative multi-agent networks, where each agent has the same goal, such as in consensus networks \cite{kashyap07}, diversification in smart systems brings about inevitable mismatches in the objectives of different agents. This then leads to noncooperative game formulations for smart systems in the disclosure of information \cite{akyol, CDC15, crawford1982strategic,tamura2014theory}. As an example, a trajectory controller can drive a tracking system to a desired path, different from the tracker's actual intent, by controlling the disclosed information \cite{sayinCDC16}.

To this end, consider the scenario of a sender (S) having access to some information and a receiver (R) needing this information to be able to take a particular action, impacting both S and R. In the classical communication setting, S seeks to transmit this information in the best possible way, leading to a full cooperation between him/her and R, toward mitigating the channel's impact on the transmitted signals. However, even if there exists an ideal (perfect) channel between S and R, if their objectives differ, absolute transparency of the disclosed information is not a reasonable action for S in general \cite{akyol, CDC15, crawford1982strategic,tamura2014theory}. In a hierarchical game, also known as Stackelberg game, \cite{BasarBook}, R reacts after S's disclosure of information. Therefore, in strategic settings, where objectives differ, S develops strategies\footnote{In this paper, we use the terms ``policy", ``strategy", and ``signaling rule" interchangeably.} to control the transparency of the disclosed information.

Originally, a scheme of the type introduced in the previous paragraph, called strategic information transmission, was introduced in a seminal paper by V. Crawford and J. Sobel \cite{crawford1982strategic}, and attracted significant attention in the economics literature due to the wide range of relevant applications, from advertising to expert advise sharing problems \cite{cheapTalk, battaglini02, ottaviani06}. Here, in addition to the privately accessed information, S's objective function includes a bias term, while R's objective function is independent of the bias. Under Nash equilibrium, in which the players announce their strategies simultaneously, the authors have shown that a quantization-based mapping of the information achieves Nash equilibrium \cite{crawford1982strategic}. In \cite{saritacs2015quadratic}, the authors have analyzed the approach of \cite{crawford1982strategic} in noisy and multidimensional settings for quadratic cost functions. In \cite{dynSIT}, the authors have extended the one-shot game of strategic information transmission to a multistage one with a finite horizon, such that the information provider and decision makers interact several times regarding a constant unknown state of the world.

Recently, strategic information transmission in hierarchical signaling games (where there is a hierarchy in the announcement of the strategies) has attracted substantial interest in various disciplines, including control theory \cite{sayinCDC16,farokhi2014gaussian,saritas16ISIT}, information theory \cite{akyol,CDC15}, and economics \cite{tamura2014theory, gentzkow2011bayesian}. In \cite{farokhi2014gaussian}, the authors have studied strategic sensor networks for Gaussian variables and with myopic quadratic objective functions, i.e., the players construct strategies just for the current stage irrespective of the length of the horizon, by restricting the receiver strategies to affine functions. Reference \cite{sayinCDC16} has addressed the optimality of linear sender strategies within the general class of policies for myopic quadratic objectives. In \cite{saritacs2015quadratic, saritas16ISIT}, the authors have shown that for scalar parameters, quadratic cost functions, and a commonly known bias parameter, the hierarchical game formulation can be converted into a team problem. Reference \cite{akyol} has shown that linear sender strategies achieve the equilibrium within the general class of policies even with additive Gaussian noise channels. In \cite{tamura2014theory}, the author has demonstrated the optimality of linear sender strategies also for the multivariate Gaussian information, and with quadratic cost functions. In \cite{gentzkow2011bayesian}, the authors have addressed the optimality of full or no disclosure for general information parameters.

In addition to the mismatched objectives in a communication system, the signaling game setting can also be considered as a dynamic deception game \cite{greenberg82,greenberg82opr, li09}, where a player aims to deceive the other player, say victim, such that the victim's perception about an underlying phenomenon and correspondingly the victim's reaction is controlled in a desirable way. Hence, this approach brings about new security and resilience applications for cyber-physical systems that are vulnerable to cyber attacks, e.g., power grids, transportation systems, and cloud networks \cite{zhu15,han15dependable,han15info}. In particular, turning the problem around, new defense mechanisms can be developed aiming to arouse attackers' suspicion on compromised information or to deceive attackers to take certain actions. Additionally, the resulting scheme would be advantageous for the defender, i.e., sender in the strategic communication scheme, in terms of his/her objectives due to the hierarchical structure, and therefore would be more preferable for security related scenarios.

Recent Verizon Data Breach Investigation Report (DBIR) \cite{DBIR} shows that millions of people have been affected by and substantial amount of financial loss has occurred due to cyber attacks. Furthermore, many of the attacks are either unreported or not yet discovered by the victim. Importantly, in $93\%$ of the attacks, the attackers can infiltrate into the system within minutes and even seconds, and in $68\%$ of the attacks the attackers exfiltrate the system within days \cite{DBIR}. Therefore, if attackers succeed in infiltrating into the system, an additional layer of defense based on dynamic deception can play a vital role for the security of the system \cite{cyberDeception}. The experiment conducted in blue (defender) and red (attacker) teams from Lockheed Martin \cite{lockheedMartin} is an illustrative example of the effectiveness of deception strategies even when the confidential information has been compromised.

Now, coming to the specifics of this paper, we obtain here equilibrium achieving sender strategies in hierarchical (i.e., Stackelberg \cite{BasarBook}) multistage signaling games with finite horizon, where hierarchically S is the leader such that his/her strategies are known by (and enforced on) R. We show that memoryless linear sender strategies and linear receiver strategies can yield multistage equilibrium with finite horizon for general quadratic objective functions and multivariate Gaussian processes evolving according to first order auto-regressive models. This extends the result for the optimality of linear strategies shown in \cite{tamura2014theory} to the dynamic settings. We point out that in the dynamic settings, in addition to the mismatches between the objectives, S should also control the transparency of the disclosed information due to impact of the actions on future stages. At each stage, S faces a trade-off in terms of the current stage and all other future stages of the game while controlling the transparency of the disclosed information, and should develop strategies in a comprehensive manner over the horizon. We point out that reference \cite{sayinCCA16} addresses, in a two-stage setting, the purity of sender strategies (whether policies should include irrelevant information or not) by restricting the sender policies to affine functions, but does not completely solve the problem. However, here we show that pure linear strategies achieve the equilibrium within the general class of policies.

After obtaining equilibrium achieving policies in the multistage strategic communication game, we extend the results to noncooperative strategic control, where sensor and controller of a dynamic system have different objectives. As an example, the controller aims to drive the system to a desired path based on the sensor outputs, while the sensor designs the sensor outputs to deceive the controller so that the system is driven to a path different from the controller's actual intent. Such a scheme can have important applications in resilience of cyber-physical systems under adversarial attacks. Even though attackers have infiltrated into the controller and gained access to control the system, the damage could be minimized via the strategic sensor outputs. {\cb Furthermore, the sensors of the system could also be infiltrated into by the attackers, which can annihilate the proposed defense mechanism via a shortcut to the state realization if the sensors's policies could be controlled remotely. In order to mitigate that, we consider the scenario where the sensor's signaling rules are {\em selected beforehand} to minimize the expected loss and {\em fixed} (can be time-variant, yet not controlled) during the operation. We provide an algorithm to compute the optimal linear sensor signaling rules for Gauss-Markov processes controlled by the controller with any measurable control rule\footnote{For linear signaling rules, the optimal control rules are also linear \cite{stochasticbook}.} numerically with global optimality guarantees.}

Particularly, the proposed formulation can be considered as a passive defense strategy that can be incorporated along with active defense strategies \cite{cyberDeception}. Consider the scenarios, where infiltration detection mechanisms (an active defense mechanism) have detected adversarial infiltration into the controller of the cyber-physical system and characterized the control objective of the adversary. And there is certain necessary time before disabling the access of the attacker to the controller. For that time interval, which can be considered as the time horizon in our formulation, the system can switch to the proposed passive defense mode, where the sensor outputs have been constructed to minimize the damage due to the attack. We emphasize that no information will be shared with the attacker except the sensor's outputs (which can also be non-informative). Importantly, {\em the defender will not provide his/her strategy as to how the sensor's outputs are constructed to the attacker.} However, the problem still can be considered as a Stackelberg game, where the sensor is the leader, because this is a passive defense mechanism that does not depend on the actual realizations, i.e., the sensor seeks to minimize the expected cost, and the attacker, having access to the system, can be aware of the switch to this passive defense mode and correspondingly can know how the sensor outputs will have been constructed.

The main contributions {\cb and conclusions} of this paper are as follows:
\begin{itemize}
  \item We study dynamic hierarchical Gaussian signaling games with finite horizon for general quadratic cost functions {\cb and general class of measurable policies}.
  \item We formulate a functional minimization problem whose solutions correspond to the equilibrium achieving signaling rules, and we characterize the solutions through a finite dimensional optimization problem bounding the original problem from below.
  \item We show that {\em linear} sender and receiver signaling rules can yield the equilibrium for arbitrary (finitely many) number of stages {\cb within the general class of measurable policies}.
  \item We show that in multistage case, the sub-games at each-stage are not decoupled and cannot be considered as single-stage games, where the innovation part of the state is to be disclosed.
  \item {\cb We geometrically observe that linear strategies for Gaussian information can achieve the equilibrium within the general class of measurable policies because uncorrelatedness implies independence.
  \item Correspondingly, the proposed method, to characterize the solution via lower bound, may not compute the optimal strategies for arbitrary distributions in general unless, e.g., the receivers' strategies are restricted to linear functions.
  \item We also argue that Gaussian distribution is the {\em best} distribution serving the deceptive sender's objectives.} 
  \item {\cb We extend the results to controlled Gauss-Markov processes and provide an algorithm to compute optimal linear sensor strategies numerically with global optimality guarantees in noncooperative control scenarios.}
\end{itemize}

The paper is organized as follows: In Section \ref{sec:problem}, we provide the problem description. In Section \ref{sec:disclosure}, we introduce and analyze strategic communication scenario, and we provide the equilibrium achieving policies in Section \ref{sec:equilibrium}. {\cb In Section \ref{sec:intriguing}, we highlight intriguing properties of hierarchical signaling games.} We analyze {\cb strategic communication in noncooperative control scenarios} in Section \ref{sec:control}. We provide numerical examples for different noncooperative communication and control scenarios in Section \ref{sec:example}. We conclude the paper in Section \ref{sec:conclusion} with several remarks. Appendices provide proofs for technical results.

{\bf Notations:}  For an ordered set of parameters, e.g., $x_1,\cdots,x_n$, we define $x_{[k,l]} := x_k,\cdots,x_l$, where $1\leq k \leq l \leq n$. ${\mathbb N}(0,.)$ denotes the multivariate Gaussian distribution with zero mean and designated covariance. We denote random variables by bold lower case letters, e.g., $\rx$. For a random variable $\rx$, $\hx$ is another random variable corresponding to its posterior belief conditioned on certain random variables that will be apparent from the context. For a vector $x$ and a matrix $A$, $x'$ and $A'$ denote their transposes, and $\|x\|$ denotes the Euclidean ($L^2$) norm of the vector $x$. For a matrix $A$, $\mathrm{tr}\{A\}$ denotes its trace. We denote the identity and zero matrices with the associated dimensions by $I$ and $O$, respectively. For positive semi-definite matrices $A$ and $B$, $A\succeq B$ means that $A-B$ is also a positive semi-definite matrix.

\section{Problem Description}\label{sec:problem}
Consider a controlled stochastic system described by the following state equation\footnote{The provided derivations can be extended to time-variant cases rather routinely. And the derivations can also be extended to the non-zero mean case in a straight-forward way.}:
\begin{align}
&\rx_{k+1} = A \rx_k + B \ru_k + \rw_k,\label{eq:sis}
\end{align}
for $k=1,\ldots,n$, where\footnote{We assume that the matrix $A$ is non-singular.} $A \in \mathbb{R}^{p\times p}$, $B\in\mathbb{R}^{p\times t}$, $\rx_1 \sim \mathbb{N}(0,\Sigma_1)$. The additive noise process $\{\rw_k\}$ is a white Gaussian vector process, e.g., $\rw_k \sim \mathbb{N}(0,\Sigma_w)$, and is independent of the initial state $\rx_1$. The closed loop control vector $\ru_k \in \mathbb{R}^t$ is given by
\begin{equation}
\ru_k = \gamma_k(\ry_{[1,k]}),\label{eq:u_def}
\end{equation}
where $\gamma_k(\cdot)$ is a Borel measurable function from $\mathbb{R}^{pk}$ to $\mathbb{R}^{t}$. The message signal $\ry_k \in \mathbb{R}^{p}$ is given by
\begin{equation}
\ry_k  = \eta_k(\rx_{[1,k]}),\label{eq:y_def}
\end{equation}
where $\eta_k(\cdot)$ is a Borel measurable function from $\mathbb{R}^{pk}$ to $\mathbb{R}^p$. We assume that the auto-covariance matrices $\Sigma_1$ and $\Sigma_w$ are all positive definite.

\begin{figure}[t!]
  \centering
  \includegraphics[width=3in]{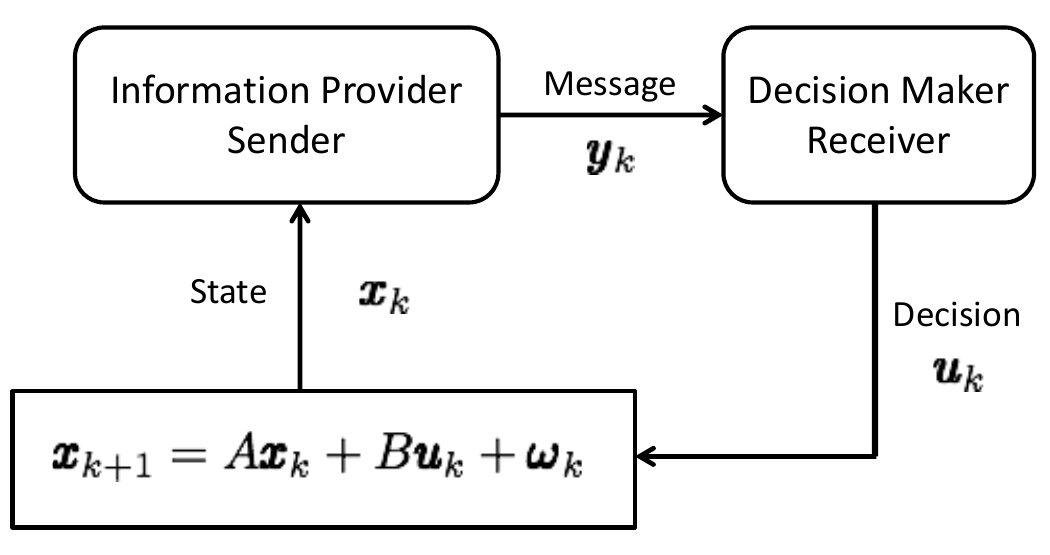}
  \caption{multistage signaling game model.}\label{fig:model}
\end{figure}

As seen in Fig. \ref{fig:model}, we have two agents: Sender (S) and Receiver (R), who select signaling rules under different objectives. For stage $k$, S selects the signaling rule $\eta_k(\cdot)$ from the policy space $\Omega_k$, which is the set of all Borel measurable functions from $\mathbb{R}^{kp}$ to $\mathbb{R}^{p}$, i.e., $\eta_k\in\Omega_k$, such that $\ry_k = \eta_k(\rx_{[1,k]})$ {\cb almost everywhere over $\mathbb{R}^p$}. On the other side, R selects the signaling rule $\gamma_k(\cdot)$ from the policy space $\Gamma_k$, which is the set of all Borel measurable functions from $\mathbb{R}^{kp}$ to $\mathbb{R}^t$, i.e., $\gamma_k \in \Gamma_k$, such that $\ru_k = \gamma_k(\ry_{[1,k]})$ {\cb almost everywhere over $\mathbb{R}^t$}. S and R have different quadratic finite horizon cost functions\footnote{We provide these functions explicitly in the following sections in {\cb noncooperative communication and control scenarios}.} $J_S(\eta_{[1,n]};\gamma_{[1,n]})$ and $J_R(\eta_{[1,n]};\gamma_{[1,n]})$, respectively, while each signaling rule implicitly depends on the other. In the following, we introduce a hierarchical equilibrium concept for the signaling rules with respect to these cost functions, $J_S$ and $J_R$. Particularly, we consider the situation where there is a hierarchy between the agents in the announcement of the policies such that S leads the game by announcing and sticking to his/her policies beforehand and R reacts to those policies accordingly. We can model such a scheme as a Stackelberg game between the players \cite{BasarBook} such that the leader, i.e., S, chooses his signaling rule based on the corresponding best response of the follower, i.e., R.

Due to the hierarchy, R's signaling rule $\gamma_k$ can depend on S's signaling rules $\eta_{[1,k]}$. In order to explicitly show the dependence on S's policies, henceforth, we denote R's policies by\footnote{Without loss of generality, we can also consider that $\gamma_k(\eta_{[1,n]}) = \gamma_k(\eta_{[1,k]})$.} $\gamma_k(\eta_{[1,k]})$, i.e., $\gamma_k(\eta_{[1,k]})(\ry_k) := \gamma_k(\ry_{[1,k]})$. Then, for each $n$-tuple of policies $\eta_k \in \Omega_k$, $k=1,\ldots,n$, we let $\React_R(\eta_{[1,n]})$ be the reaction set of R, as a subset of $\bigtimes_{k=1}^n \Gamma_k$. For finite-horizon objectives, we have
\[
\React_R(\eta_{[1,n]}) := \arg \min\limits_{\substack{\gamma_k\in\Gamma_k,\\ k=1,\ldots,n}}J_R(\eta_{[1,n]};\gamma_{[1,n]}(\eta_{[1,n]})),
\]
where 
\(
\gamma_{[1,n]}(\eta_{[1,n]}) := \{\gamma_1(\eta_1),\ldots,\gamma_n(\eta_{[1,n]})\}.
\)
In the following sections, when we provide the objective functions for the associated scenarios explicitly, we will also show that $\React_R$ is an equivalence class such that all $\gamma^*_{[1,n]}\in\React_R$ lead to the same random variable $\ru_k^*$ almost surely under certain convexity assumptions. Therefore, the pair of signaling rules $\left(\eta^*_{[1,n]},\gamma^*_{[1,n]}\right)$ attains the multistage Stackelberg equilibrium provided that
\boxalign[0.45\textwidth]{
\begin{subequations}\label{eq:eq}
\begin{align}
&\eta^*_{[1,n]} = \arg\min\limits_{\substack{\eta_k\in\Omega_k,\\ k=1,\ldots,n}} J_S\big(\eta_{[1,n]};\gamma^*_{[1,n]}(\eta_{[1,n]})\big),\\
&\gamma_{[1,n]}^*(\eta_{[1,n]})= \arg \min\limits_{\substack{\gamma_k\in\Gamma_k,\\ k=1,\ldots,n}}J_R(\eta_{[1,n]};\gamma_{[1,n]}(\eta_{[1,n]})).
\end{align}
\end{subequations}
}

\begin{rem}
We note that the hierarchical equilibrium \eqref{eq:eq} implies that S's signaling rules do not depend on the realizations of the random variables and all of them can be selected beforehand since they also do not depend on R's signaling rules. Such an equilibrium formulation, where S strategies do not depend on the actual realizations, i.e., where S does not have access to the realizations, is essential for the cyber-security related applications in order to avoid shortcuts that can cancel the proposed defense mechanism once the attacker also infiltrates into S. Furthermore, S should anticipate R's reaction to any selected strategy since even if S might have incentive to come up with another policy based on R's policy, any change in S's policy would also imply a change in R's policy accordingly due to the hierarchy.  
\end{rem}

In the following sections, we analyze the equilibrium achieving signaling rules in {\cb noncooperative communication and control scenarios}.

\section{Strategic Information Disclosure}\label{sec:disclosure}

Here, we consider a strategic communication scenario, which is a special case of \eqref{eq:sis}, where $B = O$ such that state $\{\rx_k\}$ is an exogenous process rather than a controlled process. Following this, we use the results obtained to characterize the equilibrium achieving {\cb linear} signaling rules for the original case \eqref{eq:sis}, i.e., in {\cb noncooperative} control scenario. The underlying state, now, is a Markov process (not necessarily stationary) evolving according to first-order auto-regressive model:
\begin{equation}
\rx_{k+1} = A \rx_k + \rw_k, \;\; k=1,\ldots,n\label{eq:auto}
\end{equation}
and at stage-$k$, state $\rx_k$ is a zero-mean Gaussian random vector with auto-covariance matrix $\Sigma_k$, which is given by the following recursion: $\Sigma_k = A\Sigma_{k-1}A' + \Sigma_w$ for $k=2,\ldots,n$. Note that if S and R have the same cost functions, the best signaling rule of S can be direct information disclosure, i.e., $\eta_k(\rx_{[1,k]}) = \rx_k$ {\cb almost everywhere over $\mathbb{R}^p$}, since there is a perfect channel between the agents. However, when S and R's cost functions are different, direct information disclosure may not be in S's best interest. 

Consider the situation where S and R have the following quadratic finite horizon cost functions, {\cb for $j=S,R$, respectively:
\begin{align}
J_j(\eta_{[1,n]};\gamma_{[1,n]}) = \E\left\{\sum_{k=1}^n \|\Q_{j,k} \rx_k + \R_{j,k} \ru_k\|^2\right\},\label{eq:S_cost}
\end{align}
where $\Q_{j,k} \in \mathbb{R}^{r\times p}$ and $\R_{j,k}\in\mathbb{R}^{r \times t}$}. Note that $\ru_k = \gamma_{k}(\eta_{[1,k]})(\ry_{[1,k]})$ while $\ry_l = \eta_l(\rx_{[1,l]})$ almost surely. We assume that $\R_{R,k}'R_{R,k}^{}$ is positive definite, i.e., $\R_{R,k}$ is full rank,  for all $k=1,\ldots,n$. Then, there is a linear relationship between the best R policies $\gamma_k^*$ and the posterior 
\[
\hx_k := \E\{\rx_k | \ry_{[1,k]}\}
\]
{\cb almost everywhere over $\mathbb{R}^p$}. Since $B=O$, by \eqref{eq:S_cost} for $j=R$, $\gamma_k$ only impacts the sub-cost function at stage-$k$, i.e., $\E\{ \|\Q_{R,k} \rx_k + \R_{R,k} \ru_k\|^2\}$. Correspondingly, given $\eta_{[1,k]}$, the best R strategy is given by
\[
\gamma_k^*(\eta_{[1,k]}) = \arg\min_{\gamma_k\in\Gamma_k} \E\left\{\|\Q_{R,k} \rx_k + \R_{R,k} \gamma_k(\eta_{[1,k]})(\ry_{[1,k]})\|^2\right\}
\]
and by the positive definiteness assumption on $\R_{R,k}' \R_{R,k}^{}\succ O$, we have
\begin{equation}
\gamma_k^*(\eta_{[1,k]})(\ry_{[1,k]}) = -(\R_{R,k}'\R_{R,k}^{})^{-1}\R_{R,k}'\Q_{R,k}^{} \,\hx_k,\nn
\end{equation}
almost everywhere {\cb over $\mathbb{R}^p$}, which also implies that $\React_R(\eta_{[1,n]})$ is a singleton. 

\begin{exmp}\label{ex:bias}
This noncooperative communication formulation between S and R also covers the schemes where there exist two separate exogenous processes such that, e.g., \cite{sayinCDC16},
\begin{equation}
\underbrace{\begin{bmatrix} \rz_{k+1} \\ \rtheta_{k+1} \end{bmatrix}}_{=\rx_{k+1}} = \underbrace{\begin{bmatrix} A_z & \\ & A_{\theta} \end{bmatrix}}_{= A} \underbrace{\begin{bmatrix} \rz_{k} \\ \rtheta_{k} \end{bmatrix}}_{=\rx_k} + \underbrace{\begin{bmatrix} \romega_k \\ \rnu_k \end{bmatrix}}_{=\rw_k}.\label{eq:special}
\end{equation}
R aims to track the process $\{\rz_k\}$ through the disclosed information $\ry_k$ while S wants R's decision $\ru_k$ to track a linear combination of $\rz_k$ and a bias parameter $\rtheta_k$, e.g., $\rz_k + D_k \rtheta_k$. In particular, the cost functions are in this case given by
\begin{align}
&J_S(\eta_{[1,n]};\gamma_{[1,n]}) = \E\left\{\sum_{k=1}^n \|\rz_k + D_k \rtheta_k - \ru_k\|^2\right\}\label{eq:lossS}\\
&J_R(\eta_{[1,n]};\gamma_{[1,n]}) = \E\left\{\sum_{k=1}^n \|\rz_k - \ru_k\|^2\right\}\label{eq:lossR}
\end{align}
such that in {\cb \eqref{eq:S_cost} for $j=S,R$}, $\Q_{S,k} = \begin{psmallmatrix} I & D_k \end{psmallmatrix}$, $\Q_{R,k} = \begin{psmallmatrix} I & O\end{psmallmatrix}$, and $\R_{S,k} = \R_{R,k} = -I$ for $k=1,\ldots,n$.
\end{exmp}

We note that in \cite{saritas16ISIT}, the authors also study dynamic signaling games, but for a scalar state and a commonly known bias parameter in the setup of Example \ref{ex:bias}. Therefore, different from our scheme, in \cite{saritas16ISIT} the noncooperative Stackelberg game turns into a strategically equivalent team problem such that S and R end up having the same objectives.

Corresponding to R's best reactions, S seeks policies $\eta_k^*$ which minimize the expected cost over $\eta_k\in \Omega_k$. Particularly, the optimization problem faced by S is given by
\begin{equation}
\min\limits_{\substack{\eta_k\in\Omega_k,\\ k=1,\ldots,n}} \sum_{k=1}^n\E\left\{\|\Q_{S,k} \rx_k - \R_{S,k}(\R_{R,k}'\R_{R,k}^{})^{-1}\R_{R,k}'\Q_{R,k}^{}\,\hx_k\|^2\right\}.\label{eq:fin}
\end{equation}
Note that \eqref{eq:fin} is a functional optimization problem, where S seeks to find $n$ functions among all Borel measurable functions from $\mathbb{R}^{kp}$ to $\mathbb{R}^p$, for $k=1,\ldots,n$. In order to find these functions, we first aim to formulate an optimization problem over finite dimensional spaces that bounds the original functional optimization problem \eqref{eq:fin} from below. The objective in \eqref{eq:fin} is a quadratic function of $\rx_k$ and $\hx_k$, and in the following we show that the cost function \eqref{eq:fin} can be written in terms of the second-order moments of $\rx_k$ and $\hx_k$:
\begin{align}
\E&\big\{\|\Q_{S,k} \rx_k  -  \R_{S,k}(\R_{R,k}'\R_{R,k}^{})^{-1}\R_{R,k}'\Q_{R,k}^{}\,\hx_k\|^2\big\}\nn\\
&= \E\Big\{\rx_k' \Q_{S,k}'\Q_{S,k}^{}\rx_k\Big\} - 2\E\Big\{\hx_k'\Lambda_k'\Q_{S,k}\rx_k\Big\}+\E\Big\{\hx_k'\Lambda_k'\Lambda_k \hx_k\Big\},\nn
\end{align}
where $\Lambda_k:= \R_{S,k}(\R_{R,k}'\R_{R,k}^{})^{-1}\R_{R,k}'\Q_{R,k}^{}$. Note that the first term on the right hand side does not include $\hx_k$, and therefore does not depend on S's signaling rules. For the second term, we have
\begin{align}
\E\{\hx_k'\Delta\rx_k\} &\stackrel{(a)}{=}\E\{\E\{\hx_k'\Delta\rx_k|\ry_{[1,k]}\}\} \nn\\
&\stackrel{(b)}{=} \E\{\hx_k'\Delta\E\{\rx_k|\ry_{[1,k]}\}\} \stackrel{(c)}{=} \E\{\hx_k'\Delta\hx_k\},\label{eq:exp}
\end{align}
where $\Delta$ is an arbitrary deterministic matrix with associated dimensions. The equality $(a)$ is due to the law of iterated expectations; $(b)$ holds because $\hx_k$ is $\sigma$-$\ry_{[1,k]}$ measurable; and $(c)$ is due to $\hx_k = \E\{\rx_k|\ry_{[1,k]}\}$. Therefore,
\begin{align}
2\E\left\{\hx_k'\Lambda_k'\Q_{S,k}^{}\rx_k\right\} &= \E\left\{\hx_k'(\Lambda_k'\Q_{S,k}^{}+\Q_{S,k}'\Lambda_k^{})\rx_k\right\}\nn\\
&= \E\left\{\hx_k'(\Lambda_k'\Q_{S,k}^{}+\Q_{S,k}'\Lambda_k^{})\hx_k\right\}.\label{eq:nn}
\end{align}
Then, we can re-write the optimization problem \eqref{eq:fin} as
\begin{equation}
\min\limits_{\substack{\eta_k\in\Omega_k,\\ k=1,\ldots,n}} \sum_{k=1}^n \E\{\hx_k'V_k^{}\hx_k^{}\},\label{eq:main}
\end{equation}
where
\begin{equation}\label{eq:vk}
V_k := \Lambda_k'\Lambda_k^{} - \Lambda_k'\Q_{S,k}^{} - \Q_{S,k}'\Lambda_k^{}.
\end{equation}
As an example, for Example \ref{ex:bias}, we have $V_k := \begin{psmallmatrix} -I & -D_{k}\\ -D_{k}& O\end{psmallmatrix}$.

We point out that in Reference \cite{tamura2014theory}, the author addresses multidimensional information disclosure for the single-stage case. To this end, he constructs a Semi-Definite Programming (SDP) problem as a bound on sender's objective function (named utility function in \cite{tamura2014theory}) and shows that linear strategies for Gaussian parameters can achieve this bound. Here, we employ a similar approach to extend these results to the dynamic settings by addressing the question of whether the linear strategies are still optimal within the general class of policies or not.

The first-order moment of $\hx_k$ is 
\begin{align}
\E\{\hx_k\} &= \E\{\E\{\rx_k|\ry_{[1,k]}\}\} \nn\\
&= \E\{\rx_k\} = 0
\end{align}
by the law of iterated expectations. We define the covariance matrix of $\hx_k$, namely the posterior covariance, as $\H_k := \E\{(\hx_k - \E\{\hx_k\})(\hx_k - \E\{\hx_k\})'\} = \E\{\hx_k^{}\hx_k'\}$. We note that for multivariate Gaussian variables, the mean is well defined, which implies that $\hx_k$ exists by the Radon-Nikodym theorem \cite{billingsley2008probability}. Furthermore, being multivariate Gaussian, the state parameter is integrable,  i.e., $\E\{|\rx_k|\} < \infty$, hence $\hx_k$ is finite almost surely, which implies that $\H=\E\{\hx_k^{}\hx_k'\}$ also exists.

The following lemma provides a lower bound for the minimization problem in \eqref{eq:main}.

\begin{lem}\label{lem:prob}
There exists a semi-definite programming (SDP) problem bounding the minimization problem \eqref{eq:main} from below and given by\footnote{$\mathbb{S}^p$ denotes the set of symmetric $p \times p$ matrices.}
\begin{align}\label{eq:upp}
\min\limits_{\substack{S_k \in \mathbb{S}^{p},\\ k=1,\ldots,n}}\sum_{k=1}^n\mathrm{tr}\{V_kS_k\}
\end{align}
subject to $\Sigma_j \succeq S_j \succeq AS_{j-1}A'$ for $j=1,\ldots,n$, and $S_0 = O$.
\end{lem}
\vspace{-.2in}
\begin{pf}
The proof is provided in Appendix \ref{pf:prob}. {\cb The key point is that at stage-$k$, the covariance of the posterior, i.e., $H_k$, is bounded from above by the case when we disclose the information fully, i.e., $\Sigma_k$, and is bounded from below by the case when we disclose no information yet R can still infer $\rx_k$ based on all previously sent signals $\ry_{[1,k-1]}$.} \hfill $\square$
\end{pf}
\vspace{-.2in}
We point out that \eqref{eq:upp} is indeed an SDP problem \cite{SDP}. There exist effective computational tools to solve SDP problems numerically, e.g., through CVX, a package for specifying and solving convex programs \cite{cvx,gb08}. However, closed-form solutions can rarely be obtained \cite{SDP}. Therefore, in order to solve \eqref{eq:upp} analytically, we develop a different approach and characterize the solutions without computing them explicitly. 
The following theorem characterizes the solution of \eqref{eq:upp} for an arbitrary (but finite) number of stages.

\begin{thm} \label{thm:neces}
There exist symmetric idempotent matrices $P_k \in \mathbb{S}^{p}$, for $k=1,\ldots,n$, such that
\begin{align}
S_k^* = AS_{k-1}^*A' + (\Sigma_k - AS_{k-1}^*A')^{1/2}P_k(\Sigma_k - AS_{k-1}^*A')^{1/2},\label{eq:s2}
\end{align}
for $k=1,\ldots,n$ (with $S_0^* = O$), attains the global minimum of \eqref{eq:upp}.
\end{thm}
\vspace{-.2in}
\begin{pf} We first point out that the constraint set in \eqref{eq:upp}, i.e.,
\begin{align}
\Psi := \left\{(S_1,...,S_n)\in\mathbb{S}^{p}\times...\times\mathbb{S}^{p} \Big| \bigwedge_{k} \{S_k \in \Psi_{k}(S_{k-1})\}\right\},\label{eq:const}
\end{align}
where $\Psi_k(S_{k-1}) := \{S_k \in \mathbb{S}^{p} \;|\; \Sigma_k \succeq S_k \succeq AS_{k-1}A'\}$, is convex. To show this, consider $n$-tuples of symmetric matrices $(M_1,\ldots,M_n) \in \mathbb{S}^{p}\times\cdots\times\mathbb{S}^{p}$ and $(N_1,\ldots,N_n) \in \mathbb{S}^{p}\times\cdots\times\mathbb{S}^{p}$ such that both $(M_1,\ldots,M_n)$ and $(N_1,\ldots,N_n)$ are in the constraint set $\Psi$.
Then, $\Psi$ is a convex set if, and only if, for any $t \in [0,1]$, the linear combination
\begin{align}
(E_1,\ldots,E_n) :&= t(M_1,\ldots,M_n) + (1-t)(N_1,\ldots,N_n) \nn\\
&= (tM_1+(1-t)N_1,\ldots,tM_n+(1-t)N_n) \in \Psi.\nn
\end{align}
Since $\Sigma_1 \succeq M_1 \succeq O$ and $\Sigma_1 \succeq N_1 \succeq O$, we have
\[
\Sigma_1 \succeq tM_1 + (1-t)N_1 \succeq O,
\]
and $E_1 = tM_1 + (1-t)N_1 \in \Psi_1(O)$. Suppose that $E_j\in\Psi_j(E_{j-1})$ for $j=1,\ldots,k-1$. Since $\Sigma_k \succeq M_k \succeq AM_{k-1}A'$, $\Sigma_k \succeq N_k \succeq AN_{k-1}A'$, and $E_{k-1} = tM_{k-1}+(1-t)N_{k-1}$, we obtain
\[
\Sigma_k \succeq tM_k + (1-t)N_k \succeq AE_{k-1}A',
\]
and $E_k = tM_k + (1-t)N_k \in \Psi_k(E_{k-1})$. By induction, we conclude that the convex combination $(E_1,\ldots,E_n) \in \Psi$ and therefore $\Psi$ is a convex set. Note that since the objective function in \eqref{eq:upp} is linear in $S_1,\ldots,S_n$ (not a zero function) and the constraint set is non-empty compact (since $\Psi$ is a cartesian product of the closed and bounded sets $\Psi_k(S_{k-1})$) and convex, the global minimum can be attained at the {\em extreme points} of $\Psi$.\footnote{An extreme point of a convex set is a point that cannot be written as a convex combination of any other points in the interior of the set.}

Next, we formulate the extreme points of $\Psi$. To this end, for given $S_{-k} := \{S_1,\ldots,S_{k-1},S_{k+1},\ldots,S_n\}$, we introduce the sub-constraint set:
\begin{align}
\Phi_{k}(S_{-k}) := \big\{S_k \in \mathbb{S}^{p} \;| \;&\Sigma_k \succeq S_k \succeq AS_{k-1}A' \nn\\
\wedge \;A^{-1}S_{k+1}&(A')^{-1} \succeq S_k \succeq A S_{k-1}A' \big\},\label{eq:sub}
\end{align}
for each $k=1,\ldots,n$, where we set $S_0 = O$ and $S_{n+1} = A\Sigma_{n}A' + \Sigma_w$. We consider that the sub-constraint set $\Phi_k(S_{-k})=\varnothing$ is empty if $\Sigma_k - AS_{k-1}A'$ or $A^{-1}S_{k+1}(A')^{-1} - AS_{k-1}A'$ are not positive semi-definite. Then, the following lemma provides a necessary condition for the extreme points of $\Psi$ in terms of these sub-constraint sets \eqref{eq:sub}.

\begin{lem} \label{lem:neces}
If an $n$-tuple $(E_1,\ldots,E_n) \in \Psi$ is an extreme point of $\Psi$, then for each $k=1,\ldots,n$, $E_k \in \Phi_{k}(E_{-k})$ is an extreme point of $\Phi_k(E_{-k})$.
\end{lem}
\vspace{-.2in}
\begin{pf}
The proof {\cb can be shown via contradiction and} is provided in Appendix \ref{pf:neces}. \hfill $\blacksquare$
\end{pf}
\vspace{-.2in}
Now, we seek to obtain the extreme points through the necessary conditions provided in Lemma \ref{lem:neces}. Let $(S_1^*,\ldots,S_n^*)\in\Psi$ be an extreme point of $\Psi$. Then, the element $S_n^*$ should be an extreme point of $\Phi_n(S_{-n}^*)$. To this end, consider arbitrary $S_1,\ldots,S_n$. Then, in stage-$n$, the sub-constraint set $\Phi_n(S_{-n})$ is given by
\[
\Phi_n(S_{-n}) = \{S_n \in \mathbb{S}^{p}| \Sigma_n \succeq S_{n} \succeq AS_{n-1}A'\}
\]
since we set $S_{n+1} = A\Sigma_nA' + \Sigma_w$ and $A^{-1}(A\Sigma_nA' + \Sigma_w) (A')^{-1} = \Sigma_n + A^{-1}\Sigma_w(A')^{-1}\succ \Sigma_n$.
We note that for each $k=1,\ldots,n$, if $\Sigma_k \succeq S_{k}$, then the matrix $\Sigma_{k+1} - AS_{k}A'$ is positive definite because
\[
\Sigma_{k+1} - AS_kA' = A\Sigma_k A' + \Sigma_w - AS_kA'=A(\Sigma_k - S_k)A' + \Sigma_w
\]
and $\Sigma_w \succ O$ by definition. Then, if $\Sigma_{n-1} \succeq S_{n-1}$, we have $\Sigma_n \succ AS_{n-1}A'$ and the following transformation:
\begin{align}
F_n(S_n) := (\Sigma_n - AS_{n-1}A')^{-1/2}(S_n-&AS_{n-1}A')\nn\\
&\times(\Sigma_n - AS_{n-1}A')^{-1/2}\nn
\end{align}
such that $F_n$ maps the sub-constraint set $\Phi_n(S_{-n})$ to
\[
F_n(\Phi_n(S_{-n})) = \{P\in\mathbb{S}^{p}| I \succeq P \succeq O\}.
\]
The following lemma characterizes the extreme points of the convex set $\Phi:=\{P \in \mathbb{S}^{p} \;|\; I\succeq P\succeq O\}$.

\begin{lem}\label{lem:idem}
A point $P_e$ in $\Phi$ is an extreme point if, and only if, $P_e$ is a symmetric idempotent matrix.
\end{lem}
\vspace{-.2in}
\begin{pf}
The proof {\cb can be shown via contradiction and} is provided in Appendix \ref{pf:idem}. \hfill $\blacksquare$
\end{pf}
\vspace{-.2in}
We note that under  bijective affine transformation of a convex set, the extreme points are mapped to the extreme points of the transformed set \cite{sierksma}. Since $F_n(\cdot)$ is a bijective affine transformation, $P_o \in \Phi$ is an extreme point of $\Phi$ if, and only if, $F^{-1}_n(P_o) \in \Phi_n(S_{-n})$ is an extreme point of $\Phi_n(S_{-n})$. Therefore, if $\Sigma_{n-1} \succeq S_{n-1}$, the extreme points of $\Phi_n(S_{-n})$ are given by
\begin{equation}
S_n^* = AS_{n-1}A' + (\Sigma_n - AS_{n-1}A')^{1/2}P_n(\Sigma_n - AS_{n-1}A')^{1/2},\nn
\end{equation}
where $P_n$ is a symmetric idempotent matrix.

For stage-$(n-1)$, we have the sub-constraint set:
\begin{align}
\Phi_{n-1}(S_{-(n-1)}) = \{S_{n-1} \in \mathbb{S}^{p}| \Sigma_{n-1} &\succeq S_{n-1} \succeq AS_{n-2}A'\nn\\
\wedge \;A^{-1}S_n(A')^{-1}&\succeq S_{n-1} \succeq AS_{n-2}A'\}.\nn
\end{align}
We point out that if $S_{n-1}\in\Phi_{n-1}(S_{-(n-1)})$, we have $\Sigma_{n-1} \succeq S_{n-1}$. Then, setting $S_n = S_n^*$, we obtain
\begin{align}
\Phi_{n-1}(S_{-(n-1)}) = \{S_{n-1} \in \mathbb{S}^{p}| \Sigma_{n-1} &\succeq S_{n-1} \succeq AS_{n-2}A'\nn\\
\wedge \;S_{n-1} &+ \Delta \succeq S_{n-1} \succeq AS_{n-2}A'\},\nn
\end{align}
where
\[
\Delta := A^{-1}(\Sigma_n - AS_{n-1}A')^{1/2}P_n(\Sigma_n - AS_{n-1}A')^{1/2}(A')^{-1} \succeq O.
\]
Therefore, if $S_n$ is an extreme point of $\Phi_n(S_{-n})$, the sub-constraint set $\Phi_{n-1}(S_{-(n-1)})$ can be written as
\[
\Phi_{n-1}(S_{-(n-1)}) = \{S_{n-1} \in \mathbb{S}^{p}| \Sigma_{n-1} \succeq S_{n-1} \succeq AS_{n-2}A'\}.
\]
Correspondingly, if $\Sigma_{n-2} \succeq S_{n-2}$, the extreme points of $\Phi_{n-1}(S_{-(n-1)})$ are given by 
\[
S_{n-1}^* = AS_{n-2}A' + (\Sigma_{n-1} - AS_{n-2}A')^{1/2}P_{n-1}(\Sigma_{n-1} - AS_{n-2}A')^{1/2},
\]
where $P_{n-1}$ is also a symmetric idempotent matrix. Since $\Sigma_{n-1} \succeq S_{n-1}^* $, setting $S_{n-1}^{} = S_{n-1}^*$, we have
\[
S_n^* = AS_{n-1}^*A' + (\Sigma_n - AS_{n-1}^*A')^{1/2}P_n(\Sigma_n - AS_{n-1}^*A')^{1/2}.
\]
Following identical steps, we obtain that any extreme point $(S_1^*,\ldots,S_n^*)$ of $\Psi$ should satisfy \eqref{eq:s2}.\hfill $\square$
\end{pf}
\vspace{-.2in}
In the next section, we address the tightness of the bound \eqref{eq:upp}, i.e., whether {\cb the minimum of} the lower bound can be achieved through certain sender policies or not.

\section{{\cb Optimality of Linear Sender} Signaling Rules}\label{sec:equilibrium}

Even though Theorem \ref{thm:neces} characterizes the necessary and sufficient conditions for the minimizing arguments of the SDP problem \eqref{eq:upp}, it still does not provide the solutions explicitly. However, as we will show next, these results have important consequences in the characterization of equilibrium achieving signaling rules for the original optimization problem \eqref{eq:main}. In particular, sender strategies that can be constructed to yield posterior covariances in \eqref{eq:s2} can minimize the lower bound \eqref{eq:upp}, and therefore can minimize the main objective function \eqref{eq:main}.

The following theorem says that for {\em any} solution of \eqref{eq:upp}, say $S_1^*,\ldots,S_n^*$, there exist certain deterministic matrices $\B_k \in \mathbb{R}^{p\times p}$ for $k=1,\ldots,n$, such that the memoryless linear disclosure policies
\begin{align}
\eta_k(\rx_{[1,k]}) = \B_k'\rx_k,\label{eq:lin}
\end{align}
{\cb almost everywhere over $\mathbb{R}^p$}, result in $\H_1=S_1^*,\ldots,\H_n=S_n^*$. In particular, by minimizing the lower bound on S's objective function, the memoryless linear sender policies \eqref{eq:lin} yield the multistage Stackelberg equilibrium within the general class of {\cb measurable} policies.

\begin{thm}\label{thm:equ}
Let $S_1^*,\ldots,S_n^*$ be a solution of the SDP problem \eqref{eq:upp} and $P_1,\ldots,P_n$ be the corresponding symmetric idempotent matrices in \eqref{eq:s2}. Let $P_k$, $k=1,\ldots,n$, have the eigen decompositions: $P_k = U_k \Lambda_k U_k'$. Then, for
\begin{equation}\label{eq:bk}
\B_k = (\Sigma_k - AS_{k-1}^*A')^{-1/2}U_k\Lambda_k,
\end{equation}
memoryless linear sender strategies \eqref{eq:lin} yield the multistage equilibrium \eqref{eq:main} within the general class of policies.
\end{thm}
\vspace{-.2in}
\begin{pf}
Say that S employs memoryless linear policies as in \eqref{eq:lin} for some deterministic combination of matrices $\B_1,\ldots,\B_n \in \mathbb{R}^{p\times p}$. Correspondingly, the posteriors $\hx_k = \E\{\rx_k|\ry_{[1,k]}\} = \E\{\rx_k|\B_1'\rx_1^{},\ldots,\B_k'\rx_k^{}\}$ are given by\footnote{We take the pseudo inverse of the matrices since they can be singular if the associated matrix $\B_k$ has a rank smaller than $p$. For example, at stage-$k$, S can disclose no information $\eta_{k}(\rx_{[1,k]}) = 0$.}
\begin{align}
\hx_1 = &\Sigma_1 \B_1 (\B_1' \Sigma_1 \B_1)^{\dagger} \B_1' \rx_1\label{eq:hats1}\\
\hx_k = &A\hx_{k-1} + (\Sigma_k - A\H_{k-1}A')\B_k(\B_k'(\Sigma_k - A\H_{k-1}A')\B_k)^{\dagger}\nn\\
&\times\B_k'(\rx_k - A\hx_{k-1}) \mbox{ for } k \geq 2.\label{eq:hatsk}
\end{align}
Next, we seek to compute $\H_k = \E\{\hx_k\hx_k'\}$, for $k=1,\ldots,n$. By \eqref{eq:hats1}, we obtain
\(\H_1 = \Sigma_1 \B_1 (\B_1'\Sigma_1^{} \B_1^{})^{\dagger} \B_1' \Sigma_1\)
since $\E\{\rx_1^{}\rx_1'\} = \Sigma_1$, and for a matrix $M$ and its pseudo-inverse $M^{\dagger}$, we have $M^{\dagger}MM^{\dagger} = M$. By \eqref{eq:hatsk}, for $\H_2$, we have a cross-term $\E\{\hx_1(\rx_2 - A\hx_1)'\}$, which can be written as
\begin{align}
\E\{\hx_1(\rx_2 - A\hx_1)'\} &= \E\{\hx_1^{}\rx_2'\} - \E\{\hx_1^{}\hx_1'\}A'\nn\\
&= \H_1 A' - \H_1 A' = O, \nn
\end{align}
due to the law of iterated expectations such that $\E\{\hx_1^{}\rx_2'\} = \E\{\E\{\hx_1^{}\rx_2'|\ry_1^{}\}\} = \E\{\hx_1^{}\E\{\rx_2'|\ry_1^{}\}\} = \E\{\hx_1^{}\hx_1'\}A'$.
Then, \eqref{eq:hatsk}, for $k\geq2$, leads to
\begin{align}
\H_k = \;&A\H_{k-1}A' + (\Sigma_k - A\H_{k-1}A')\B_k\nn\\
&\times (\B_k'(\Sigma_k - A\H_{k-1}A')\B_k)^{\dagger} \B_k'(\Sigma_k - A\H_{k-1}A').\nn
\end{align}
Let $\C_1 := \Sigma_1^{1/2}\B_1$ and $\C_k := (\Sigma_k - A\H_{k-1}A')^{1/2}\B_k$ for $k=2,\ldots,n$, such that $\H_1 = \Sigma_1^{1/2}\C_1^{}(\C_1'\C_1^{})^{\dagger}\C_1'\Sigma_1^{1/2}$, and for $k\geq 2$,
\begin{align}
\H_k = A\H_{k-1}A' + &(\Sigma_k - A\H_{k-1}A')^{1/2}\nn\\
&\times \C_k^{}(\C_k'\C_k^{})^{\dagger}\C_k'(\Sigma_k - A\H_{k-1}A')^{1/2}.\label{eq:HH}
\end{align}
Note that $\C_k^{}(\C_k'\C_k^{})^{\dagger}\C_k'$, for $k=1,\ldots,n$, is a symmetric idempotent matrix and the posterior covariances $\H_1,\ldots,\H_n$ have identical expressions as in \eqref{eq:s2}.
If the symmetric idempotent matrices $P_k$ for $k=1,\ldots,n$ corresponding to the minimizers of the SDP problem \eqref{eq:upp} have the eigen decompositions: $P_k = U_k\Lambda_kU_k'$, we can set $\C_k = U_k \Lambda_k$ for $k=1,\ldots,n$, such that $\C_k^{}(\C_k'\C_k^{})^{\dagger}\C_k' = P_k$. In particular, setting $\B_1 = \Sigma_1^{-1/2}U_1\Lambda_1$ and $\B_k = (\Sigma_k - AS^*_{k-1}A')^{-1/2}U_k\Lambda_k$, we obtain $\H_k=S_k^*$ for $k=1,\ldots,n$. Hence, the memoryless linear signaling rules \eqref{eq:lin} can minimize the main objective function \eqref{eq:main} within the general class of {\cb measurable} policies. \hfill $\square$
\end{pf}
\vspace{-.2in}
In Table \ref{tab:desc}, we provide a description to compute the equilibrium achieving sender policies based on Lemma \ref{lem:prob}, and Theorems \ref{thm:neces} and \ref{thm:equ}.
We note that for linear sender signaling rules, the corresponding equilibrium achieving receiver signaling rules are also linear since the underlying state is Gaussian. Therefore, linear sender and receiver signaling rules can achieve the equilibrium {\cb also in multistage hierarchical Gaussian signaling games}.

\begin{table}[t!]
\renewcommand{\arraystretch}{1.5}
  \caption{A description to compute equilibrium achieving sender policies in strategic communication.\\}
  \centering
  \begin{tabularx}{.47\textwidth}{ l}
          \hline
          \textbf{Algorithm \ref{tab:desc}:} Strategic Communication \\
          \hline
          \textbf{SDP Problem:} \\
          \hspace{1em} {\em Compute $V_k$, $\forall k$, by \eqref{eq:vk}.}\\
          \hspace{1em} {\em Solve the SDP problem \eqref{eq:upp} through a numerical toolbox}\\
          \hspace{2em} {\em and obtain the solution $S_k^*$, $\forall k$.}\\
          \textbf{Equilibrium achieving policies:}\\
          \hspace{1em} {\em Compute the corresponding idempotent matrices $P_k$,$\forall k$,}\\
          \hspace{2em} {\em by using the solution $S_k^*$, $\forall k$, and \eqref{eq:s2}.}\\
          \hspace{1em} {\em Compute the eigen decompositions: $P_k = U_k\Lambda_kU_k'$.}\\
          \hspace{1em} {\em Compute $\B_k$, $\forall k$, by using $S_{k-1}^*,U_k,\Lambda_k$, and \eqref{eq:bk}.}\\
          \hline
  \end{tabularx}\label{tab:desc}
\end{table}

\section{Intriguing Properties of Hierarchical Signaling}\label{sec:intriguing}

In this section, we list several intriguing remarks related to the proposed hierarchical signaling scheme:

{\cb \bf Uniqueness of the solution:} In addition to the memoryless linear sender strategies in Theorem \ref{thm:equ}, any signaling rule leading to the same posteriors can yield the equilibrium. As an example, if $\ry_k = \B_k' \rx_k$ for $k=1,\ldots,n$ achieves the equilibrium, then $\tilde{\ry}_1 = \ry_1$ and $\tilde{\ry}_k = \ry_k - \E\{\ry_k|\ry_{[1,k-1]}\}$ for $k=2,\ldots,n$ lead to the same posteriors, i.e., $\E\{\rx_k|\tilde{\ry}_{[1,k]}\} = \E\{\rx_k|\ry_{[1,k]}\}$ for $k=1,\ldots,n$, therefore yield the equilibrium. Note that $\tilde{\ry}_{[1,n]}$ is a whitened version of $\ry_{[1,n]}$, i.e., $\tilde{\ry}_k$'s are pair-wise independent of each other.

{\cb \bf Inter-stage Coupling:} In general, when S and R have different cost functions, signaling rules: $\eta_k(\rx_{[1,k]}) = \BB_k' (\rx_k-\E\{\rx_k|\rx_{[1,k-1]}\})$, for certain matrices $\BB_k \in \mathbb{R}^{p\times p}$ (where $\rx_k-\E\{\rx_k|\rx_{[1,k-1]}\} = \rw_{k-1}$ is the innovation in the state process) do not lead to the equilibrium, contrary to the case when they have the same cost functions. In particular, in the multistage case, the sub-games at each stage are not decoupled and cannot be considered as a single stage game as if the innovation part of the state is going to be disclosed. As an example, let $\Sigma_1 = O$ such that $\rx_2 = \rw_1$; then $\ry_2 = \BB_2'\rw_1$ and $\ry_3 = \BB_3'\rw_2$. This implies that 
\begin{align}
&\H_2 = \Sigma_w\BB_2(\BB_2'\Sigma_w \BB_2)^{\dagger}\BB_2'\Sigma_w\nn\\
&\H_3 = A\H_2A' + \Sigma_w \BB_3 (\BB_3' \Sigma_w \BB_3)^{\dagger}\BB_3' \Sigma_w.\nn
\end{align} 
However, by Theorem \ref{thm:neces}, the corresponding $S_k$'s are
\begin{align}
&S_2 = \Sigma_2^{1/2}P_2 \Sigma_2^{1/2} = \Sigma_w^{1/2}P_2\Sigma_w^{1/2},\nn\\
&S_3 = AS_2A' + (\Sigma_3 - AS_2A')^{1/2}P_3(\Sigma_3 - AS_2A')^{1/2}.\nn
\end{align}
We can set $\BB_2$ such that $\H_2 = S_2$; however, $\H_3 = S_3$ requires that
\begin{align}
\Sigma_w^{1/2} \BB_3 (\BB_3' \Sigma_w \BB_3)^{\dagger}&\BB_3' \Sigma_w^{1/2}= \Sigma_w^{-1/2}(\Sigma_3 - AS_2A')^{1/2}\nn\\
&\times P_3(\Sigma_3 - AS_2A')^{1/2}\Sigma_w^{-1/2}.\label{eq:req}
\end{align}
Note that the left hand side of \eqref{eq:req} is a symmetric idempotent matrix, however, the right hand side is not necessarily an idempotent matrix. 

{\cb \bf Applicability to Noisy Observations:} The results would hold if the message space was larger than $p$ since it would also lead to the same constraint set $\Psi$ \eqref{eq:const}. However, the derivations would not carry over if S had access to noisy version of the state instead of the actual state. As an example, consider the situation where S has access to $\rs_k = C\rx_k + \rv_k$, where $C\in\mathbb{R}^{p \times p}$ and $\{\rv_k\sim \mathbb{N}(0,\Sigma_v)\}$ is an independent white Gaussian noise process. Then, for $\ry_k = \B_k' \rs_k$, the posteriors would be given by
\begin{align}
&\hx_1 = \Sigma_1C' \B_1(\B_1'C\Sigma_1C'\B_1 + \underbrace{\B_1' \Sigma_v \B_1})^{\dagger}\B_1' \rs_1\nn\\
&\hx_k = A\hx_{k-1} + (\Sigma_k-A\H_{k-1}A')C'\B_k\nn\\
&\times\big[\B_k'C(\Sigma_k-A\H_{k-1}A')C'\B_k + \underbrace{\B_k' \Sigma_v\B_k}\big]^{\dagger}\B_k' (\rs_k - CA\hx_{k-1})\nn
\end{align}
and due to the underbraced term, $(\Sigma_k - A\H_{k-1}A')^{-1/2}(\H_k - A\H_{k-1}A')(\Sigma_k - A\H_{k-1}A')^{-1/2}$ would not lead to a symmetric idempotent matrix, contrary to \eqref{eq:HH}.

{\cb \bf Applicability to Different Dimensional Signals:} For the single stage case, i.e., $n=1$, as in \cite{tamura2014theory}, the lower bound \eqref{eq:upp} is given by
\begin{equation}
\min_{S\in\mathbb{S}^p}\, \mathrm{tr}\{V_1S\}
\end{equation}
subject to $\Sigma_1 \succeq S \succeq O$. And the optimizer is given by 
\begin{equation}\label{eq:lowsol}
S^* = \Sigma_1^{1/2}Q_-Q_-'\Sigma_1^{1/2},
\end{equation} 
where $Q_- = \begin{psmallmatrix} q_1 & \cdots & q_m \end{psmallmatrix}$ and $q_j\in\mathbb{R}^p$, $j=1,\ldots,m$, are the eigenvectors of $\Sigma_1^{1/2}V_1\Sigma_1^{1/2}$ corresponding to negative eigenvalues. This also implies that in the multistage case, i.e., $n>1$, the rank of $\B_k$ is bounded from above by the number of negative eigenvalues of $V_k$ due to Sylvester's law of inertia \cite{hornma}.

{\cb 
\begin{figure}[t]
  \centering
  \includegraphics[width=1.5in]{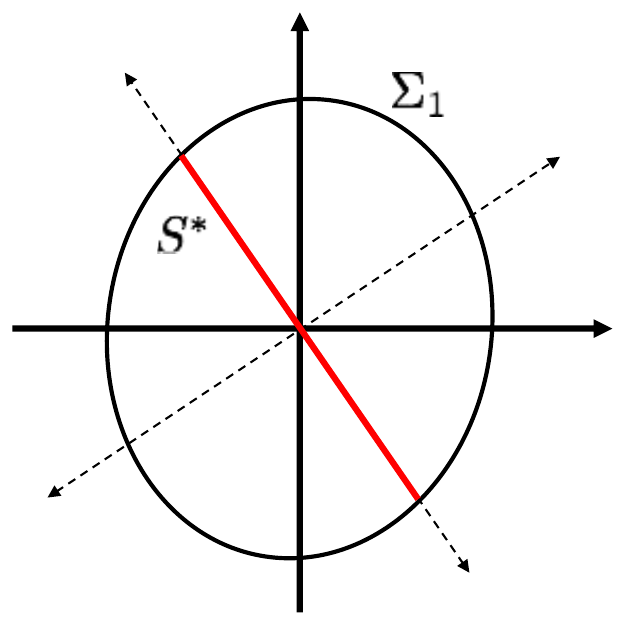}
  \caption{\cb Illustration of $\Sigma_1$ and $S^*$ for $n=2$.}\label{fig:rank}
\end{figure}

{\bf Applicability to Different Information Models:} For the single stage case, the optimal signaling rule can be computed analytically and is given by \cite{tamura2014theory}
\begin{equation}\label{eq:singleopt}
\eta(\rx_1) = Q'\Sigma_1^{-1/2}\rx_1,
\end{equation}
almost everywhere over $\mathbb{R}^p$, where $Q = \begin{psmallmatrix} Q_- & O_{p\times (p-m)}\end{psmallmatrix}$. When we take a closer look at \eqref{eq:lowsol} for $p=2$, we observe that the signal should not be informative in a certain direction (e.g., the direction of the eigenvector of $\Sigma_1^{1/2}V_1\Sigma_1^{1/2}$ associated with positive eigenvalue) while it should be fully informative another direction (e.g., the direction of the eigenvector associated with negative eigenvalue), which has also been illustrated in Fig. \ref{fig:rank}. Then, the optimal signaling rule for Gaussian information case, i.e., \eqref{eq:singleopt}, is just a projection of the information onto the direction (expected to be fully informative) through a linear signaling rule. Since the signals in the orthogonal directions\footnote{Note that they have zero-mean at those directions.} are uncorrelated with each other and they are jointly Gaussian, the orthogonality implies that they are independent of each other. Correspondingly, any information on one of them, e.g., vertical direction, does not provide any information about the other, e.g., horizontal direction, since the posterior conditioned on independent information is just the same with the posterior without conditioning on any information. However, this is not the case for arbitrary information models, e.g., other than Gaussian, since uncorrelatedness does not imply independence in general.  

\begin{figure}[t]
  \centering
  \includegraphics[width=3.3in]{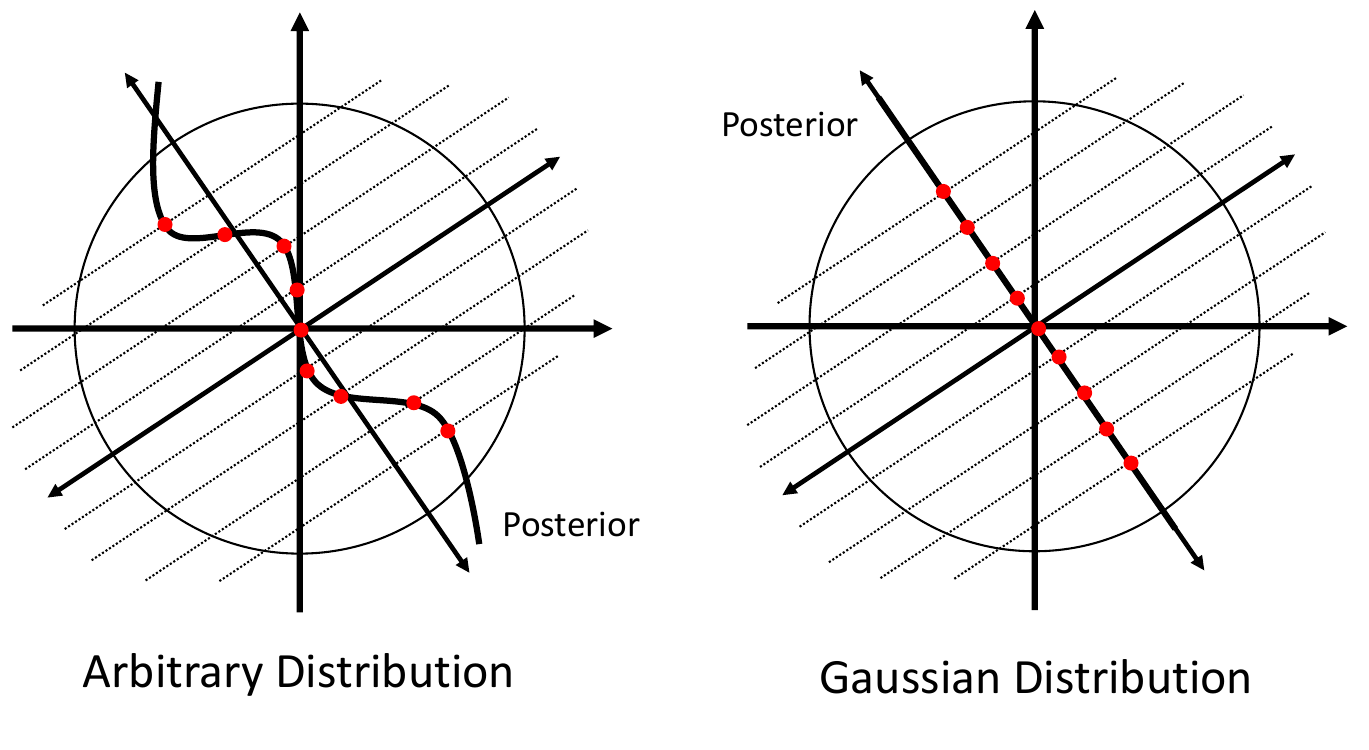}
  \caption{\cb For $p=2$, we show how informative the signal is in the other direction if the signal is the projection of the realization on the left-tilted direction for arbitrary distribution (on the left sub-figure) and Gaussian distribution (on the right sub-figure) while $\Sigma_1 = I$. Particularly, the projection implies that the realization could be at any point on the line passing through the projection, and parallel to the right-tilted direction. Conditioned on that, the posterior (marked with red dots) may not be on the left-tilted direction, i.e., the projection can also be informative on the right-tilted direction for arbitrary distributions.}\label{fig:means}
\end{figure}

In particular, Fig. \ref{fig:means} illustrates this for $p=2$ and when the underlying coordinates, i.e., $\mathbb{R}^p$, have been transformed linearly such that $\Sigma_1=I$ over the new coordinates. We seek to design the signaling rule such that the signal is fully informative in the left-tilted direction while not informative in the right-tilted direction. The projection onto the left-tilted direction is fully informative in that direction. However, such a projection implies that the realized information can be at any point on the line passing through the projection, and parallel to the right-tilted direction according to the underlying distribution. For the Gaussian case, the corresponding posterior is just on the left-tilted direction since $\Sigma_1 = I$. However, for arbitrary distributions, the posterior conditioned on the parallel lines may not be on the left-tilted direction, as seen in Fig. \ref{fig:means}, and any deviation of the posterior from the left-tilted direction implies that the projection-based signaling is also informative on the right-tilted direction. Therefore, the covariance of the posterior, i.e., $\E\{\hx_1^{}\hx_1'\}$, would not be equal to the solution of the lower bound \eqref{eq:lowsol} through such a linear (or indeed any) signaling rule for arbitrary distributions in general.

{\bf Optimality of Gaussian Information for Deception:} What we have observed above also implies that Gaussian distribution is the {\em best} distribution serving deceptive S's objective since the lower bound is not necessarily tight for other distributions. Additionally, if R's policies are restricted to linear functions, then the proposed approach also computes the optimal signaling rule for arbitrary distributions, since uncorrelatedness yields linear independence. }

In the next section, we {\cb compute the optimal linear signaling rules in noncooperative control scenarios}, e.g., $B\neq O$ in \eqref{eq:sis}.

\section{\cb Hierarchical Signaling in Noncooperative Control Systems} \label{sec:control}

Returning back to the general scenario \eqref{eq:sis}, we now consider the situation where S and R have the following quadratic finite horizon cost functions, for $j=S,R$, respectively:
\begin{align}
&J_j = \E\left\{\sum_{k=1}^n \rx_{k+1}'\Q_{j,k+1}\rx_{k+1} + \ru_k'\R_{j,k}\ru_k\right\},\label{eq:S}
\end{align}
where $\Q_{j,k+1}\in\mathbb{R}^{p\times p}$ are positive semi-definite and $\R_{j,k}\in\mathbb{R}^{t\times t}$ are positive definite. 

For {\cb discrete-time linear Gaussian stochastic systems and in {\em scalar} settings}, if the measurement signals and the control inputs can be constructed within the class of general policies, and there exists an additive white Gaussian noise (AWGN) channel between the sensor and the controller, reference \cite{bansal89} has shown that in the simultaneous {\cb (or cooperative)} design, the optimal performance in terms of quadratic cost functions can be obtained through linear control inputs, and measurement signals which are linear in the innovation in the state. {\cb However, the linear measurement signals may not be optimal for the multidimensional cases in general. Furthermore, again in a cooperative setting, reference \cite{ref:Tanaka15} has formulated the joint sensor and controller design problem over a perfect channel when there exists additive (privacy-preservation induced) information-theoretic cost of communication in addition to the quadratic control cost similar to \eqref{eq:S}. And the authors have addressed the problem when the sensor's policies are restricted to a linear structure and provided an SDP-based algorithm to compute the optimal strategies \cite{ref:Tanaka15}.

\begin{rem}\label{rem:gen}
We note that this stochastic control problem entails non-classical information in the general settings, where S selects signaling rules within the general class of measurable policies, since S's signaling rules are functions of the actual state (correspondingly the control inputs) and both players can exploit this via triggering-based threat strategies in this {\em noncooperative} dynamic setting \cite{BasarBook,ref:Myerson97}. Correspondingly, optimality of linear sender strategies is still an open problem, and linearity may or may not hold within the general class of strategies.
\end{rem}

However, here, we restrict S's signaling rules to memoryless linear strategies as in \cite{ref:Tanaka15} and in the following, we provide an SDP-based efficient algorithm to compute the optimal sender strategies numerically with global optimality guarantees. Particularly, S selects $\signal_k$ from the space of all linear functions from $\mathbb{R}^p$ to $\mathbb{R}^p$, denoted by $\Omega^{\ell}$, such that the signal $\ryl = \signal_k(\rx_k)$ almost everyhere over $\mathbb{R}^p$. Correspondingly, there exist certain matrices $L_k\in\mathbb{R}^{p\times p}$ such that $\ryl_k=L_k'\rx_k$ almost everywhere over $\mathbb{R}^p$. We note that R still selects his/her strategies from $\Gamma_k$, $k=1,\ldots,n$.}

The above is a general framework for noncooperative game formulations between S and R, which permits analysis of the equilibrium achieving signaling rules in the most general form. By adjusting the matrices, it is possible to generate many different examples of stochastic control/game problems in strategic environments.

\begin{exmp}
As one example, there can be two separate controlled stochastic processes:
\[
\underbrace{\begin{bmatrix} \rz_{k+1} \\ \rtheta_{k+1} \end{bmatrix}}_{=\,\rx_{k+1}} = \underbrace{\begin{bmatrix} A_z & \\ & A_{\theta} \end{bmatrix}}_{=\, A} \underbrace{\begin{bmatrix} \rz_{k} \\ \rtheta_{k} \end{bmatrix}}_{=\,\rx_k} + \underbrace{\begin{bmatrix} B_z \\ B_{\theta} \end{bmatrix}}_{=\,B} \ru_k + \underbrace{\begin{bmatrix} \romega_k \\ \rnu_k \end{bmatrix}}_{=\,\rw_k}.
\]
R aims to drive $\rz_k$ into her desired path, but the control variable $\ru_k$ has also impact on the state $\rtheta_k$ and S designs the measurement signals so that $\rtheta_k$ is driven into his own desired path not in line with R's actual intent. In particular, in the objective functions, $\Q_{S,k} = \begin{psmallmatrix} O & O \\ O & \Q_{\theta,k}\end{psmallmatrix}$ and $\Q_{R,k} =\begin{psmallmatrix}\Q_{z,k}&O\\ O & O\end{psmallmatrix}$ and R seeks to minimize
\begin{equation}
\E\left\{\sum_{k=1}^n \rz_{k+1}'\Q_{z,k+1}\rz_{k+1} + \ru_k'\R_{R,k}\ru_k\right\},\label{eq:CZ}
\end{equation}
while S seeks to minimize
\begin{equation}
\E\left\{\sum_{k=1}^n \rtheta_{k+1}'\Q_{\theta,k+1}\rtheta_{k+1} + \ru_k'\R_{S,k}\ru_k\right\}.\nn
\end{equation}
\end{exmp}

\begin{exmp}\label{ex:biasedcontrol}
Another special case is one where while R seeks to drive $\rz_k$ into her desired path, S wants $\rz_k$ to track an exogenous process $\rtheta_k$, i.e., $B_{\theta} = O$. Then, S's cost function is given by
\begin{align}
\sum_{k=1}^n (\rz_{k+1} - D_{k+1}\rtheta_{k+1})'\Q_{\theta,k+1}(\rz_{k+1} - D_{k+1}\rtheta_{k+1}) + \ru_k'\R_{S,k}\ru_k\nn
\end{align}
and this corresponds to
\(
\Q_{S,k} = \begin{psmallmatrix} I \\ -D_k' \end{psmallmatrix} \Q_{\theta,k} \begin{psmallmatrix} I & -D_k \end{psmallmatrix}
\)
in \eqref{eq:S}, while R's cost function is given by \eqref{eq:CZ}.
\end{exmp}

{\cb
In order to address the hierarchical signaling in noncooperative control scenarios, rather routinely by completing the squares \cite{bansal89,stochasticbook}, for $j=S,R$, we can write \eqref{eq:S} as
\begin{align}
\sum_{k=1}^n \E\{\rx_{k+1}'&\Q_{j,k+1}\rx_{k+1} + \ru_k'\R_{j,k}\ru_k\} \nn\\
&= \sum_{k=1}^n \E\left\{\|\ru_k + \K_{j,k} \rx_k\|_{\Delta_{j,k}}^2\right\} + \Delta_{j,0}, \label{eq:SS}
\end{align}
where\footnote{The assumption $\R_{j,k}\succ O$ ensures that $\Delta_{j,k}$ is non-singular.} $\K_{j,k} = \Delta_{j,k}^{-1}B'\tQ_{j,k+1}A$, $\Delta_{j,k} = B'\tQ_{j,k+1}B + \R_{j,k}$,
\begin{align}
&\tQ_{j,k} = \Q_{j,k} + A'(\tQ_{j,k+1} - \tQ_{j,k+1}B\Delta_{j,k}^{-1}B'\tQ_{j,k+1})A\label{eq:Q}\\
&\tQ_{j,n+1} = \Q_{j,n+1},\;\Delta_{j,0} = \mathrm{tr}\{\tQ_{j,1}\Sigma_1\} + \sum_{k=1}^n\mathrm{tr}\{\tQ_{j,k+1}\Sigma_w\},\nn
\end{align}
and set $\Q_{j,1} = O$. Then, through a routine change of variables,
\begin{align}
&\sum_{k=1}^n\E\left\{\|\ru_k + K_{j,k} \rx_k\|_{\Delta_{j,k}}^2\right\} = \sum_{k=1}^n\E\left\{\|\ru_{j,k} + K_{j,k} \rx_k^o\|_{\Delta_{j,k}}^2\right\},\label{eq:comm}
\end{align}
where $\ru_{j,k} = \ru_k + K_{j,k} B \ru_{k-1} + \cdots + K_{j,k} A^{k-2}B\ru_1$ and
\begin{align}
&\rx_{k+1}^o = A\rx_k^o + \rw_k.\label{eq:free}
\end{align}
\vspace{-.2in}
\begin{rem}
In the cooperative settings, \eqref{eq:comm} would imply that the optimal control can be computed via the solution for the sub-cost function $\E\{\|\ru_{R,k} + K_{R,k}\rx_k^o\|_{\Delta_{R,k}}^2\}$. As an example, we would have $\ru_{R,1}^* = -K_{R,1}\E\{\rx_1^o|\ryl_1\}$ and $\ru_{R,2}^* = -K_{R,2}\E\{\rx_2^o|\ryl_1,\ryl_2\}$, and correspondingly, the optimal control inputs would be given by $\ru_1^* = \ru_{R,1}^*$ and $\ru_2^* = \ru_{R,2}^* - K_{R,1}\ru_{R,1}^*$, almost everywhere over $\mathbb{R}^t$. However, in the noncooperative settings, even though the control-free process $\{\rx_k^o\}$ is independent of how the control inputs $\ru_k$'s are constructed, the sensor outputs $\ryl_k$ depend on the state $\rx_k$ and correspondingly on the previous control inputs $\ru_{[1,k-1]}$. Therefore, while constructing the control inputs in the noncooperative settings, R should also consider their impact on the future stages, as also pointed out in Remark \ref{rem:gen}. 
\end{rem}

However, the following lemma shows that R cannot influence how S selects his/her linear signaling rules strategically. 

\begin{lem}\label{lem:new}
For a controlled Gauss-Markov process $\{\rx_k\}$ evolving according to \eqref{eq:sis} and the control-free state $\{\rx_k^o\}$ evolving according to \eqref{eq:free}, we have
\begin{equation}\label{eq:equ}
\E\{\rx_k^o|L_1'\rx_1,\ldots,L_k'\rx_k\} = \E\{\rx_k^o|L_1'\rx_1^o,\ldots,L_k'\rx_k^o\}
\end{equation}
where $L_j\in\mathbb{R}^{p\times p}$, $j=1,\ldots,k$.
\end{lem}
\vspace{-.2in}
\begin{pf}
By \eqref{eq:free}, the linear signaling rule yields 
\begin{equation}
L_k'\rx_k = L_k'\rx_k^o + \underbrace{L_k'B\ru_{k-1}+\ldots+L_k'A^{k-2}B\ru_1},
\end{equation}
almost everywhere over $\mathbb{R}^p$, where the under-braced term is $\sigma$-$\ryl_{[1,k-1]}$ measurable for all $k=1,\ldots,n$, which implies \eqref{eq:equ}. However, this would not necessarily be the case for nonlinear signaling rules in general since the previous control inputs could {\em limit} the informativeness of the current signal. \hfill $\square$
\end{pf}
\vspace{-.2in}
Based on Lemma \ref{lem:new}, the problem faced by S, i.e., \eqref{eq:comm}, is just a strategic information disclosure problem. Therefore, after some algebra, similar to the lines followed in Section \ref{sec:disclosure}, the problem can also be written as an affine function of $H_k^o := \E\{\hx_k^o(\hx_k^o)'\}$, where $\hx_k^o:=\E\{\rx_k^o|\ryl_{[1,k]}\}$, as follows:
\begin{equation}
\min_{\substack{\eta_k\in\Omega_k,\\ k=1,\ldots,n}} \sum_{k=1}^n \mathrm{tr}\{V_k^o H_k^o\} + \Xi_o,\label{eq:opt}
\end{equation}
for certain symmetric deterministic matrices $V_k^o \in \mathbb{R}^{p\times p}$, $k=1,\ldots,n$, which are given by
\begin{equation}
V_k^o := \Xi_{k,k} + \sum_{l=k+1}^n \Xi_{k,l}A^{l-k} + (A^{l-k})'\Xi_{l,k},\label{eq:vv}
\end{equation}
and $\Xi_{k,l}$ is the corresponding $p\times p$ sub-block of $\Xi$, which is given by
\begin{align}
&\Xi := T_S'\Delta_S T_S - T_S'\Delta_S K_S - K_S'\Delta_S T_S\label{eq:Xi}\\
&\Xi_o := \mathrm{tr}\{\Sigma^o K_S'\Delta_S K_S\} + \Delta_{S,0},
\end{align}
where $\Sigma^o := \E\{\rx^o (\rx^o)'\}$, $\rx^o := \begin{psmallmatrix} \rx_n^o&\cdots&\rx_1^o \end{psmallmatrix}'$, $T_S := \Phi_S\Phi_R^{-1}K_R$, and
\begin{align}
&\Phi_j :=\begin{psmallmatrix} I & K_{j,n} B & K_{j,n} A B & \cdots & K_{j,n} A^{n-2} B \\ & I & K_{j,n-1}B & \cdots & K_{j,n-1}A^{n-3}B \\ & & I & \cdots & K_{j,n-2}A^{n-4}B \\ & & & \ddots &  \\ & & & & I\end{psmallmatrix},\\
&K_j := \begin{psmallmatrix} K_{j,n} & & \\ & \vphantom{\int\limits^x}\smash{\ddots} & \\ & & K_{j,1}\end{psmallmatrix}, \Delta_j := \begin{psmallmatrix} \Delta_{j,n} & & \\ & \vphantom{\int\limits^x}\smash{\ddots} & \\ & & \Delta_{j,1}\end{psmallmatrix},\label{eq:KDelta}
\end{align}
for $j=S,R$. Then, Theorems \ref{thm:neces} and \ref{thm:equ} show that given $V_k$'s, we can compute the optimal linear signaling rules via Algorithm 1. In the following corollary of these theorems, we recap the results.

\begin{cor} \label{thm:control}
We can compute the optimal linear signaling rules in noncooperative control settings by computing \eqref{eq:vv} based on \eqref{eq:Q} and \eqref{eq:Xi}-\eqref{eq:KDelta}, and then applying Algorithm \ref{tab:desc} for \eqref{eq:opt}.
\end{cor}}

\begin{figure}[t!]
  \centering
  \begin{overpic}[width=.5\textwidth]{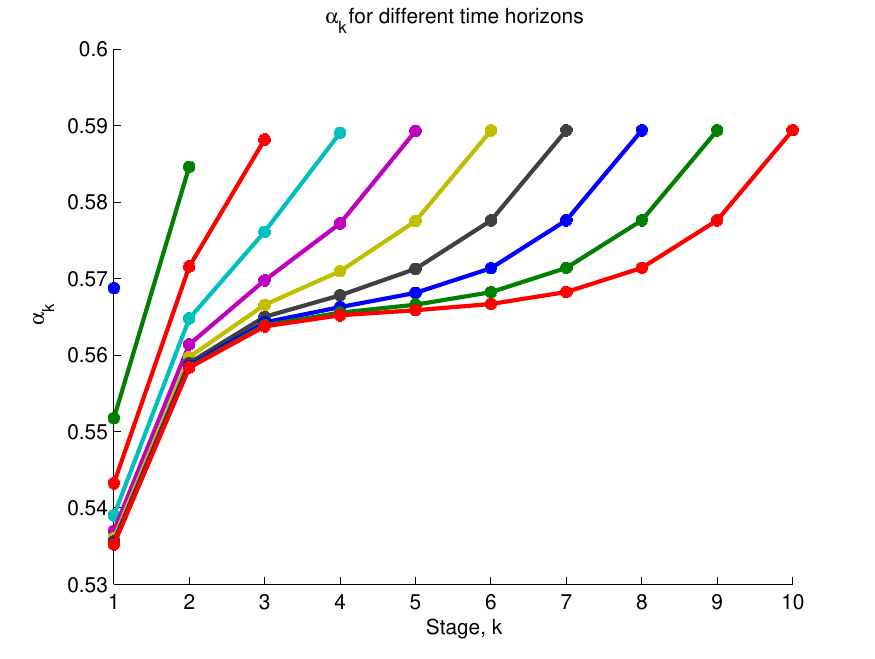}
  \put(25,21){\colorbox{white}{\parbox{0.65\linewidth}{%
     \begin{equation}\label{eqfigR}
     A = \begin{psmallmatrix} \frac{1}{\sqrt{3}}&0\\0&\frac{1}{\sqrt{2}}\end{psmallmatrix},\;B=\begin{psmallmatrix} 0 \\ 0\end{psmallmatrix},\;\Sigma_k = \begin{psmallmatrix} 3/2&0\\0&2 \end{psmallmatrix}
     \end{equation}}}}
  \end{overpic}\\
  \caption{Scenario 1: the process $\{\rz_k\}$ is relatively less colored, i.e., less correlated in time, than the process $\{\rtheta_k\}$.}\label{fig:r}
\end{figure}

\begin{figure}[t!]
  \centering
  \begin{overpic}[width=.5\textwidth]{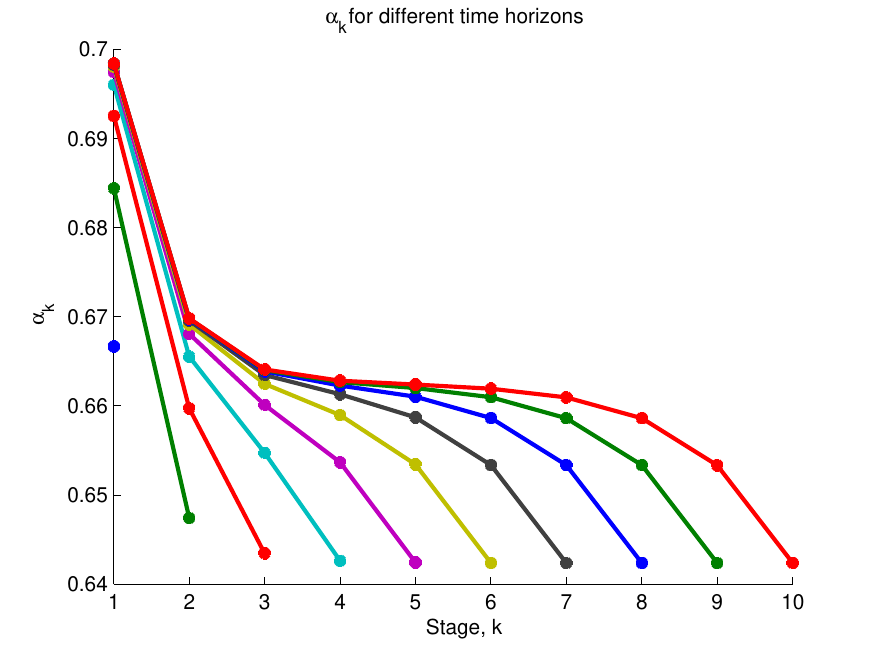}
  \put(25,55){\colorbox{white}{\parbox{0.65\linewidth}{%
     \begin{equation}\label{eqfigD}
     A = \begin{psmallmatrix} \frac{1}{\sqrt{2}}&0\\0&\frac{1}{\sqrt{3}}\end{psmallmatrix},\;B=\begin{psmallmatrix} 0 \\ 0\end{psmallmatrix},\;\Sigma_k = \begin{psmallmatrix} 2&0\\0&3/2 \end{psmallmatrix}
     \end{equation}}}}
  \end{overpic}\\
  \caption{Scenario 2: the process $\{\rz_k\}$ is relatively more colored.}\label{fig:d}
\end{figure}

\section{Illustrative Examples}\label{sec:example}

As numerical illustrations, we first consider Example \ref{ex:bias} in strategic communication formulation with cost functions \eqref{eq:lossS} and \eqref{eq:lossR}, where $\rz_k \in \mathbb{R}$, $\rtheta_k \in \mathbb{R}$, and $D_k = 1$. We consider two different scenarios: Scenario 1, where the process $\{\rz_k\}$ is relatively more colored, i.e., more correlated in time, than the process $\{\rtheta_k\}$, and Scenario 2, where the process $\{\rtheta_k\}$ is more colored. To this end, we set $\Sigma_k$ and $A$ as in \eqref{eqfigR} and \eqref{eqfigD}, introduced as parts of Figs. \ref{fig:r} and \ref{fig:d}, respectively, which yields that the underlying state process $\{\rx_k\in\mathbb{R}^2\}$ is stationary and $\Sigma_w = I$.

We can compute the equilibrium achieving sender policies via Algorithm 1. After the computation, we observe that the resulting weight matrices $\B_k\in\mathbb{R}^{2\times 2}$, $\forall k$, have rank 1, and indeed have a column that is full of zeros. Therefore, we can consider that at stage $k$, S sends practically a scalar which is a linear combination of $\rz_k$ and $\rtheta_k$ to R. Additionally, we can scale that sent signal by multiplying it with a certain constant such that the weight of $\rz_k$ in the message is just $1$. In particular, the sent signal can be written as $\ry_k = \rz_k + \alpha_k \rtheta_k$ for a certain constant $\alpha_k \in \mathbb{R}$. In Figs. \ref{fig:r} and \ref{fig:d}, we plot the time evolution of $\alpha_k$ for different time horizons, e.g., $n=1,\ldots,10$, in the Scenarios 1 and 2, respectively. We observe that the weight of $\rz_k$ and $\rtheta_k$ increases or decreases in time depending on their relative correlatedness in time. As an example, in Scenario 1, the process $\{\rtheta_k\}$ is relatively more colored than the process $\{\rz_k\}$ and the weight of $\rtheta_k$, i.e., $\alpha_k$, in the sent messages increases in time compared to the weight of $\rz_k$, i.e., $1$. Additionally, the pattern that the weight $\alpha_k$ draws as the length of time horizon grows provides an insight for the equilibrium achieving sender policies for stationary state processes in infinite time horizon, e.g., after a transient phase, the weights could reach a steady state value as $n\rightarrow \infty$.

\begin{figure}[t!]
  \centering
  \begin{overpic}[width=.5\textwidth]{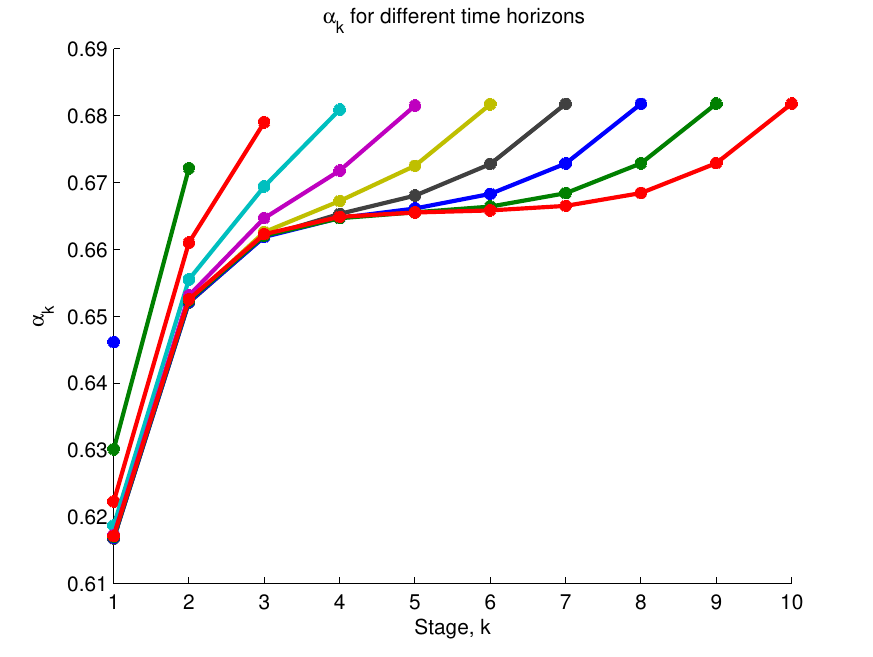}
  \put(22,25){\colorbox{white}{\parbox{0.7\linewidth}{%
     \begin{equation}\label{eqfig3}
     A = \begin{psmallmatrix} \frac{1}{\sqrt{3}}&0\\0&\frac{1}{\sqrt{2}}\end{psmallmatrix},\;B=\begin{psmallmatrix} 2 \\ -\frac{1}{2}\end{psmallmatrix},\;\Sigma_k^o = \begin{psmallmatrix} 3/2&0\\0&2 \end{psmallmatrix}
     \end{equation}}}}
  \end{overpic}\\
  \caption{Scenario 3: the process $\{\rz_k\}$ is relatively less colored in {\cb noncooperative} control.}\label{fig:3}
\end{figure}

\begin{figure}[t!]
  \centering
  \begin{overpic}[width=.5\textwidth]{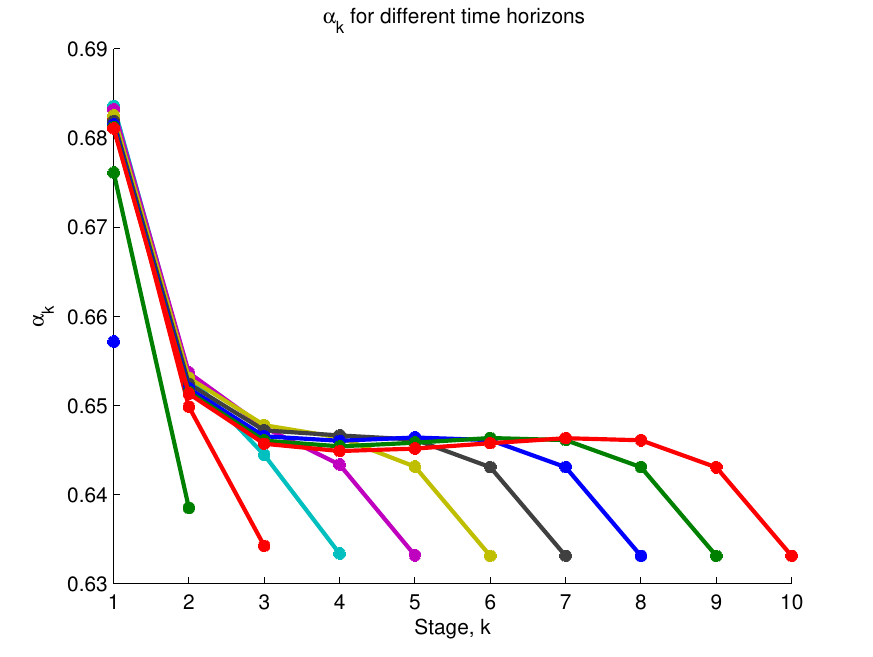}
  \put(25,55){\colorbox{white}{\parbox{0.65\linewidth}{%
     \begin{equation}\label{eqfig4}
     A = \begin{psmallmatrix} \frac{1}{\sqrt{2}}&0\\0&\frac{1}{\sqrt{3}}\end{psmallmatrix},\;B=\begin{psmallmatrix} 2 \\ -\frac{1}{2}\end{psmallmatrix},\;\Sigma_k^o = \begin{psmallmatrix} 2&0\\0&3/2 \end{psmallmatrix}
     \end{equation}}}}
  \end{overpic}\\
  \caption{Scenario 4: the process $\{\rz_k\}$ is relatively more colored in {\cb noncooperative} control.}\label{fig:4}
\end{figure}

Furthermore, we also consider Example \ref{ex:biasedcontrol} within the {\cb noncooperative} control formulation, where $\rz_k\in\mathbb{R}$, $\rtheta_k\in\mathbb{R}$, and $D_k = 1$. Different from the previous illustrative examples, here, $B = \begin{psmallmatrix} 2\\-1/2\end{psmallmatrix}$ and similar to Scenarios 1 and 2, we consider two new scenarios: Scenario 3, where the process $\{\rz_k\}$ is relatively more colored than the process $\{\rtheta_k\}$, and Scenario 4, where the process $\{\rtheta_k\}$ is more colored. To this end, we set $\Q_{\theta,k} = \Q_{z,k} = \R_{S,k} = \R_{R,k} = 1$, and $A$, $B$, and $\Sigma_k^o$ as in \eqref{eqfig3} and \eqref{eqfig4}, introduced as parts of Figs. \ref{fig:3} and \ref{fig:4}, respectively, which yields that the underlying control-free state process $\{\rx_k \in\mathbb{R}^2\}$, introduced in \eqref{eq:free}, is stationary and $\Sigma_w = I$. Similar to Scenarios 1 and 2, we also observe that all the optimal weight matrices $\B_k\in\mathbb{R}^{2\times 2}$ have rank 1 and S practically sends a scalar variable. Therefore, in Figs. \ref{fig:3} and \ref{fig:4}, we plot the change of $\alpha_k$ within various time horizons, where $\ry_k = \rz_k + \alpha_k \rtheta_k$. We also note that the evolution of $\alpha_k$ is very much also dependent on $B$ in addition to the relative colorfulness of the processes $\{\rz_k\}$ and $\{\rtheta_k\}$.

\section{Conclusion}\label{sec:conclusion}
In this paper, we have addressed the existence and characterization of the equilibrium achieving sender strategies in hierarchical signaling games with finite horizon, for quadratic objective functions and multivariate Gaussian processes. Our main conclusion has been that linear sender and receiver strategies can yield the equilibrium within the general class of policies in strategic communication scenarios. This settles an open question on the structure of equilibrium achieving policies in dynamic strategic information transmission within a Stackelberg game framework. {\cb We have observed that independence of uncorrelated Gaussian parameters plays a significant role in the optimality of linear signaling rules. Furthermore, Gaussian distribution is the best distribution serving the deceptive sender's objectives. We have also provided algorithms to compute optimal linear strategies numerically with global optimality guarantees in noncooperative communication and control scenarios.} 

Some future directions of research on this topic include characterization of the equilibrium achieving strategies in closed forms, their existence and characterization for the infinite horizon case, and {\cb analysis of equilibrium under information-regularization constraints}.

\begin{ack}                               
This research was supported in part by the U.S. Office of Naval Research (ONR) MURI grant N00014-16-1-2710, and in part by by the U.S. Army Research Labs (ARL) under IoBT Grant 479432-239012-191100.
\end{ack}

\appendix
\section{Proof of Lemma \ref{lem:prob}}\label{pf:prob}
The minimization problem in \eqref{eq:main} can be written as
\begin{equation}
\min\limits_{\substack{\eta_k\in\Omega_k,\\ k=1,\ldots,n}} \sum_{k=1}^n \mathrm{tr}\{V_k\E\{\hx_k^{} \hx_k'\}\} = \min\limits_{\substack{\eta_k\in\Omega_k,\\ k=1,\ldots,n}} \sum_{k=1}^n \mathrm{tr}\{V_k\H_k\}.\nn
\end{equation}
Note that the posterior covariances $\H_1,\ldots,\H_n$ depend on the signaling rules $\eta_1,\ldots,\eta_n$, and they are real and symmetric matrices by definition. Next, we aim to find necessary conditions on $\H_k$'s to derive a tight lower bound on \eqref{eq:main}.  To this end, consider the following positive semi-definite matrix:
\begin{align}
\E\{(\rx_k-\hx_k)(\rx_k -\hx_k)'\} = \Sigma_k - \H_k\succeq O,\nn
\end{align}
which implies $\Sigma_k \succeq \H_k$, for $k=1,\ldots,n$. Furthermore, after some algebra, it can be shown that
\begin{align}
\E\{(\hx_k-\E\{\rx_k|\ry_{[1,k-1]}\}) (\hx_k&-\E\{\rx_k|\ry_{[1,k-1]}\})'\}\nn\\
&= \H_k - A\H_{k-1}A',
\end{align}
and therefore $\H_k-A\H_{k-1}A'\succeq O$ and correspondingly $\H_k \succeq A\H_{k-1}A'$.

The posterior covariances $\H_1,\ldots,\H_k$ are real, symmetric matrices and should at least satisfy the constraints: $\Sigma_1 \succeq \H_1 \succeq O$, and $\Sigma_k \succeq \H_k \succeq A \H_{k-1}A'$, for $k =2,\ldots,n$. Based on this, we can formulate another optimization problem \eqref{eq:upp} in which the optimization arguments $S_1,\ldots,S_n \in \mathbb{S}^{p}$ are subject to the constraints in \eqref{eq:upp}. Since we have shown that those constraints are necessary (not necessarily sufficient yet), the minimization problem \eqref{eq:upp} is a lower bound on \eqref{eq:main}. Note that by the linear objective function $\sum_k \mathrm{tr}\{V_kS_k\}$ and the semi-definiteness constraints on $S_k$, \eqref{eq:upp} is an SDP problem.

\section{Proof of Lemma \ref{lem:neces}}\label{pf:neces}
Suppose that $E=(E_1,\ldots,E_n)\in\Psi$ is an extreme point of $\Psi$ and there exists an element $E_k$ such that $E_k$ is not an extreme point of $\Phi_k(E_{-k})$. Then, there exist two distinct $M,N \in \Phi_k(E_{-k})$ such that
\(
E_k = tM+(1-t)N,
\)
for some $t \in (0,1)$. Note that since $M,N \in \Phi_k(E_{-k})$, the matrices $M$ and $N$ satisfy
\begin{align}
&A^{-1}E_{k+1}(A')^{-1}\succeq M \succeq AE_{k-1}A',\nn\\
&A^{-1}E_{k+1}(A')^{-1}\succeq N \succeq AE_{k-1}A'\nn
\end{align}
by \eqref{eq:sub}. Therefore,
\begin{align}
&E_{M} := (E_1,\ldots,E_{k-1},M,E_{k+1},\ldots,E_n) \in \Psi,\nn\\
&E_{N} := (E_1,\ldots,E_{k-1},N,E_{k+1},\ldots,E_n) \in \Psi,\nn
\end{align}
and $E_{M} \neq E_{N}$ since $M\neq N$. However, we can write the extreme point $E$ as
\(
E = tE_{M} + (1-t)E_{N},
\)
for some $t\in(0,1)$ even though $E_{M},E_{N}\in \Psi$, and this leads to a contradiction. Hence, if $(E_1,\ldots,E_n)\in\Psi$ is an extreme point, the elements $E_k$ are the extreme points of the corresponding sub-constraint sets $\Phi_k(E_{-k})$.

\section{Proof of Lemma \ref{lem:idem}}\label{pf:idem}
Note that eigenvalues of a symmetric idempotent matrix are either $0$ or $1$, and suppose that for a symmetric idempotent matrix $P\in\Phi$, there exist two distinct symmetric matrices $M\in\Phi$ and $N\in\Phi$ such that $P = tM+(1-t)N$ for some $t\in(0,1)$. Let $p_1,p_0\in\mathbb{R}^{p}$ be eigenvectors of $P$ corresponding to eigenvalues $1$ and $0$, respectively. Note that since the eigenvalues of $M$ and $N$ are bounded, for any vector $p\in\mathbb{R}^{p}$, $0\leq p'M p\leq 1$ and $0\leq p'N p \leq 1$. Then, through convex combination, we have
\begin{align}
&t p_1'M p_1 + (1-t) p_1'N p_1 = p_1'P p_1 = 1,\nn\\
&t p_0'M p_0 + (1-t) p_0'N p_0 = p_0'P p_0 = 0,\nn
\end{align}
which leads to $p_1'Mp_1 = p_1'N p_1 = 1$ and $p_0'Mp_0 = p_0'Np_0 = 0$. Therefore, $p_1$ and $p_0$ are eigenvectors of $M$ and $N$. Furthermore, the eigenvalues of $M$ and $N$ corresponding to the eigenvectors $p_1$ and $p_0$ are $1$ and $0$, respectively. Since $p_1$ and $p_0$ are arbitrary eigenvectors of $P$, the matrices $M$ and $N$ have the same eigenvalues and eigenvectors with $P$, and correspondingly $P = M = N$, which, however, yields a contradiction. In view of these contradictions, we can say that a symmetric idempotent matrix is an extreme point of $\Phi$.

Lastly, we aim to show that any other matrix which is not an idempotent matrix, say $Z$, cannot be an extreme point of $\Phi$. Let $Z$ have an eigen decomposition
$Z = Q \mathrm{diag}(\lambda_1,\ldots,\lambda_{p})Q'$.
Since $Z$ is not an idempotent matrix, there exists an eigenvalue, say $\lambda_i$, which is neither $1$ nor $0$. Then, for any $t\in(0,1)$, there exist two distinct $\lambda_{i,1},\lambda_{i,2} \in [0,1]$ such that $\lambda_i = t \lambda_{i,1} + (1-t)\lambda_{i,2}$, e.g., set $\lambda_{i,1} = \lambda_i/t$ and $\lambda_{i,2} = 0$. Correspondingly, for the matrices
\begin{align}
M:= Q\, \mathrm{diag}(\ldots, \lambda_{i,1}, \ldots)\,Q' \mbox{ and } N:= Q\, \mathrm{diag}(\ldots, \lambda_{i,2}, \ldots)\,Q',\nn
\end{align}
we have $Z = tM + (1-t)N$, yet $M\neq N$, i.e., $Z$ is not an extreme point of $\Phi$.

\bibliographystyle{plain}        
\bibliography{ref}           

\begin{thebibliography}{10}

\bibitem{DBIR}
Verizon's 2016 data breach investigation report.
\newblock
  \url{http://www.verizonenterprise.com/verizon-insights-lab/dbir/2016/}.
\newblock Accessed: 2016-11-2.

\bibitem{CDC15}
E.~Akyol, C.~Langbort, and T.~Ba\c{s}ar.
\newblock Privacy constrained information processing.
\newblock In {\em Proc. 54th IEEE Conf. Decision and Control, Japan}, pages
  4511--4516, 2015.

\bibitem{akyol}
E.~Akyol, C.~Langbort, and T.~Ba\c{s}ar.
\newblock Information-theoretic approach to strategic communication as a
  hierarchical game.
\newblock {\em Proceedings of the IEEE}, 105(2):205--218, 2017.

\bibitem{BasarBook}
T.~Ba\c{s}ar and G.~J. Olsder.
\newblock {\em Dynamic Noncooperative Game Theory}.
\newblock Society for Industrial Mathematics (SIAM) Series in Classics in
  Applied Mathematics, 1999.

\bibitem{bansal89}
R.~Bansal and T.~Ba\c{s}ar.
\newblock Simultaneous design of measurement and control strategies for
  stochastic systems with feedback.
\newblock {\em Automatica}, 25(5):679--694, 1989.

\bibitem{battaglini02}
M.~Battaglini.
\newblock Multiple referrals and multidimensional cheap talk.
\newblock {\em Econometrica}, 70(4):1379--1401, 2002.

\bibitem{billingsley2008probability}
P.~Billingsley.
\newblock {\em {Probability and Measure}}.
\newblock John Wiley \& Sons Inc., 2008.

\bibitem{SDP}
G.~Blekherman, P.~A. Parrilo, and R.~R. Thomas.
\newblock {\em Semidefinite optimization and Convex Algebraic Geometry}.
\newblock Society for Industrial Mathematics (SIAM) Series on Optimization,
  2012.

\bibitem{crawford1982strategic}
V.~Crawford and J.~Sobel.
\newblock Strategic information transmission.
\newblock {\em Econometrica}, 50(6):1431--1451, 1982.

\bibitem{farokhi2014gaussian}
F.~Farokhi, A.~Teixeira, and C.~Langbort.
\newblock Estimation with strategic sensors.
\newblock {\em IEEE Trans. Automatic Control}, 62(2):724--739, 2017.

\bibitem{cheapTalk}
J.~Farrell and M.~Rabin.
\newblock Cheap talk.
\newblock {\em J. Econ. Pers.}, 10(3):103--118, 1996.

\bibitem{gentzkow2011bayesian}
M.~Gentzkow and E.~Kamenica.
\newblock Bayesian persuasion.
\newblock {\em American Economic Review}, 101(6):2590--2615, 2011.

\bibitem{dynSIT}
M.~Golosov, V.~Skreta, A.~Tsyvinski, and A.~Wilson.
\newblock Dynamic strategic information transmission.
\newblock {\em J. Economic Theory}, 151:304--341, 2014.

\bibitem{gb08}
M.~Grant and S.~Boyd.
\newblock Graph implementations for nonsmooth convex programs.
\newblock In {\em Recent Advances in Learning and Control}, pages 95--110.
  Springer-Verlag Limited, 2008.

\bibitem{cvx}
M.~Grant and S.~Boyd.
\newblock {CVX}: Matlab software for disciplined convex programming, version
  2.1.
\newblock \url{http://cvxr.com/cvx}, March 2014.

\bibitem{greenberg82opr}
I.~Greenberg.
\newblock The effect of deception on optimal decisions.
\newblock {\em Operations Research Letters}, 4, 1982.

\bibitem{greenberg82}
I.~Greenberg.
\newblock The role of deception in decision theory.
\newblock {\em J. Conflict Resolution}, 26:139--156, 1982.

\bibitem{han15dependable}
Y.~Han, J.~Chan, T.~Alpcan, and C.~Leckie.
\newblock Using virtual machine allocation policies to defend against
  co-resident attacks in cloud computing.
\newblock {\em IEEE Trans. Dependable and Secure Computing}, 14(1):95--108,
  2017.

\bibitem{han15info}
Y.~Han, J.~Chan, T.~Alpcan, C.~Leckie, and B.~I.~P. Rubinstein.
\newblock A game theoretical approach to defend against co-resident attacks in
  cloud computing: {P}reventing co-residence using semi-supervised learning.
\newblock {\em IEEE Trans. Info. Foren. Sec.}, 11(3):556--570, 2015.

\bibitem{lockheedMartin}
K.~E. Heckman, M.~J. Walsh, F.~J. Stech, T.~A. O'boyle, S.~R. Dicato, and A.~F.
  Herber.
\newblock Active cyber network defense with denial and deception.
\newblock {\em J. Computers and Security}, 37:72--77, 2013.

\bibitem{hornma}
R.A. Horn and C.R. Johnson.
\newblock {\em Matrix Analysis}.
\newblock Cambridge University Press, 1985.

\bibitem{cyberDeception}
S.~Jajodia, V.~S. Subrahmanian, and V.~Swarup.
\newblock {\em Cyber Deception: Building the Scientific Foundation}.
\newblock Springer International Publishing, Switzerland, 2016.

\bibitem{kashyap07}
A.~Kashyap, T.~Ba\c{s}ar, and R.~Srikant.
\newblock Quantized consensus.
\newblock {\em Automatica}, 43(7):1192--1203, 2007.

\bibitem{stochasticbook}
P.~R. Kumar and P.~Varaiya.
\newblock {\em Stochastic Systems}.
\newblock Prentice-Hall, 1986.

\bibitem{li09}
D.~Li and J.~B. Cruz-Jr.
\newblock Information, decision-making and deception in games.
\newblock {\em Decision Support Systems}, 47:518--527, 2009.

\bibitem{ref:Myerson97}
R.~B. Myerson.
\newblock {\em Game Theory: {A}nalysis of Conflict}.
\newblock Harvard University Press, 1997.

\bibitem{ottaviani06}
M.~Ottaviani and P.~N. Sorensen.
\newblock Professional advice.
\newblock {\em J. Econ. Theory}, 126:120--142, 2006.

\bibitem{saritas16ISIT}
S.~Sar{\i}ta{\c{s}}, S.~Y{\"u}ksel, and S.~Gezici.
\newblock Dynamic signaling games under {N}ash and {S}tackelberg equilibria.
\newblock In {\em Proc. IEEE Internati. Symp. Information Theory}, pages
  1631--1635, 2016.

\bibitem{saritacs2015quadratic}
S.~Sar{\i}ta{\c{s}}, S.~Y{\"u}ksel, and S.~Gezici.
\newblock Quadratic multi-dimensional signaling games and affine equilibria.
\newblock {\em IEEE Trans. Automatic Control}, 62(2):605--619, 2017.

\bibitem{sayinCCA16}
M.~O. Sayin, E.~Akyol, and T.~Ba\c{s}ar.
\newblock On the structure of equilibrium strategies in dynamic {G}aussian
  signaling games.
\newblock In {\em Proc. IEEE Multi--Conf. on Syst. Control, Buenos Aires},
  pages 749--754, 2016.

\bibitem{sayinCDC16}
M.~O. Sayin, E.~Akyol, and T.~Ba\c{s}ar.
\newblock Strategic control of a tracking system.
\newblock In {\em Proc. 55th IEEE Conf. Decision and Control, Las Vegas,
  Nevada}, pages 6147--6153, 2016.

\bibitem{sierksma}
G.~Sierksma, V.~Soltan, and T.~Zamfirescu.
\newblock Invariane of convex sets under linear transformations.
\newblock {\em Linear and Multilinear Algebra}, 35:37--47, 1993.

\bibitem{tamura2014theory}
W.~Tamura.
\newblock A theory of multidimensional information disclosure.
\newblock {\em Working paper, available at SSRN 1987877}, 2014.

\bibitem{ref:Tanaka15}
T.~Tanaka and H.~Sandberg.
\newblock {S}{D}{P}-based joint sensor and controller design for information
  regularized optimal {L}{Q}{G} control.
\newblock In {\em Proc. 54th IEEE Conf. Decision and Control (CDC)}, 2012.

\bibitem{zhu15}
Q.~Zhu and T.~Ba\c{s}ar.
\newblock Game-theoretic methods for robustness, security, and resilience of
  cyberphysical control systems: {G}ames-in-games principle for optimal
  cross-layer resilient control systems.
\newblock {\em IEEE Control Systems}, 35(1):46--65, 2015.

\end{thebibliography}

\end{document}